\documentstyle[floats,tighten,preprint,eqsecnum,prd,aps,epsf]{revtex}

\newcommand{\plus}{\makebox[15pt][c]{$+$}}
\newcommand{\minus}{\makebox[15pt][c]{$-$}}
\newcommand{\er}[2]
{\hskip-0.5em\raisebox{0.08em}{\scriptsize{$\;\begin{array}{@{}l@{}}
\plus\makebox[0.25em][r]{#1\hfill} \\[-0.12em]
\minus\makebox[0.25em][r]{#2\hfill} 
\end{array}$}}}
\newcommand{\err}[2]
{\hskip-0.5em\raisebox{0.08em}{\scriptsize{$\;\begin{array}{@{}l@{}}
\plus\makebox[0.50em][r]{#1\hfill} \\[-0.24em]
\minus\makebox[0.50em][r]{#2\hfill} 
\end{array}$}}}

\newcommand{\AmS}{{\protect\the\textfont2
  A\kern-.1667em\lower.5ex\hbox{M}\kern-.125emS}}
 
\newcommand{\gtap}{{\raise.3ex\hbox{$>$\kern-.75em\lower1ex\hbox{$\sim$}}}}
\newcommand{\ltap}{{\raise.3ex\hbox{$<$\kern-.75em\lower1ex\hbox{$\sim$}}}}

\newcommand{\be}{\begin{equation}}
\newcommand{\ee}{\end{equation}}
\newcommand{\bea}{\begin{eqnarray}}
\newcommand{\eea}{\end{eqnarray}}
\newcommand{\bi}{\begin{itemize}}
\newcommand{\ei}{\end{itemize}}

\begin{document}  

\preprint{\parbox[t]{15em}{\raggedleft
FERMILAB-PUB-00/346-T \\ hep-ph/0101023\\[2.0em]}}
\draft 

\title{The Semileptonic Decays $B\to\pi l\nu$ and $D\to\pi l\nu$ from 
Lattice QCD}

\author{Aida~X.~El-Khadra,$^1$ Andreas~S.~Kronfeld,$^2$ 
Paul~B.~Mackenzie,$^2$ Sin\'{e}ad~M.~Ryan$^3$ and James~N.~Simone$^2$}

\address{%
$^1$Physics Department, University of Illinois, Urbana, IL 61801, USA \\
$^2$Fermi National Accelerator Laboratory, Batavia, IL 60510, USA \\
$^3$School of Mathematics, Trinity College, Dublin 2, Ireland} 

\date{January 2, 2001}
\maketitle

\begin{abstract}
We present a lattice QCD calculation of the form factors and
differential decay rates for semileptonic decays of the heavy-light
mesons $B$ and $D$ to the final state $\pi l\nu$.
The results are obtained with three methodological improvements over
previous lattice calculations:
a matching procedure that reduces heavy-quark lattice artifacts,
the first study of lattice-spacing dependence,
and the introduction of kinematic cuts to reduce model dependence.
We show that the main systematics are controllable (within the quenched
approximation) and outline how the calculations could be improved
to aid current experiments in the determination of~$|V_{ub}|$
and~$|V_{cd}|$.
\end{abstract} 




\section{Introduction}
\label{sec:intro}

Processes involving weak decays of $B$ and $D$ mesons are of great
interest, because they yield information on the more
poorly known elements of the Cabibbo-Kobayashi-Maskawa (CKM) matrix.
Semileptonic decays have traditionally been used to determine the
CKM matrix, for example,
$V_{ud}$ (through nuclear $\beta$-decay),
$V_{us}$ ($K_{l3}$),
$V_{cb}$ ($B \to D^{(*)}l\nu$), and
$V_{ub}$ ($b \to u l \nu$) \cite{Groom:2000in}.
In the first three cases flavor symmetries (isospin, SU(3) flavor, and
heavy quark symmetry, respectively) greatly simplify one's theoretical
understanding of the hadronic transition matrix elements.
In the symmetry limit, and at zero recoil, current conservation ensures
that the matrix elements are exactly normalized.
Even when estimates of the deviations from the symmetry limit are
difficult to calculate reliably, the deviations tend to be small.
Thus, the overall theoretical uncertainty on the decay process is
under control.
Given good experimental measurements, this procedure then determines 
the associated element of the CKM matrix.

For semileptonic decays of charmed or $b$-flavored mesons into light
mesons there are no flavor symmetries to constrain the hadronic matrix
elements.
As a result, the errors on~$|V_{ub}|$ are currently dominated by
theoretical uncertainties and are not well known \cite{Groom:2000in}.
For the same reason the best value for $|V_{cd}|$, at this time,
comes from neutrino production of charm off of valence $d$ quarks
(with the cross section from perturbative QCD), rather than from the
semileptonic $D$ decays.
In this paper we take a step towards reducing the theoretical uncertainty
by using lattice QCD to calculate the form factors for the decays
$B\to\pi l\nu$ and $D\to\pi l\nu$.
Although our results are in the quenched approximation, we introduce
several methodological improvements that carry over to full QCD.
Moreover, this work is the first to study the lattice-spacing
dependence of the form factors.

There is a considerable ongoing experimental effort on this subject, 
which will lead to measurements of the differential decay rates.
For $B\to\pi l\nu$,
\begin{equation}
	\frac{d\Gamma}{dp} = \frac{G_F^2|V_{ub}|^2}{24\pi^3}
		\frac{2m_B p^4|f_+(E)|^2}{E},
	\label{eq:dgamma}
\end{equation}
where $E=p_\pi\cdot p_B/m_B$ is the energy of the pion in the rest 
frame of the $B$~meson, and $p=\sqrt{E^2-m_\pi^2}$ is the magnitude of 
the corresponding three-momentum.
($p_\pi$ and $p_B$ are four-momenta.
For $D\to\pi l\nu$, replace $V_{ub}$ with $V_{cd}$,
$m_B$ with $m_D$, and $p_B$ with $p_D$.)
The non-perturbative form factor $f_+(E)$ parametrizes the hadronic 
matrix element of the heavy-to-light transition,
\begin{equation}
	\langle\pi (p_\pi)|{\cal V}^\mu |B(p_B)\rangle =
		f_+(E) \left[p_B+p_\pi-\frac{m_B^2-m_\pi^2}{q^2}q\right]^\mu + 
		f_0(E) \frac{m_B^2-m_\pi^2}{q^2}q^\mu ,
	\label{eq:ff}
\end{equation}
where ${\cal V}^\mu$ is the charged $b\to u$ vector current,
and $q=p_B-p_\pi$ is the momentum transferred to the leptons.
For reasons that are made clear below, we prefer to consider the form
factors $f_+$ and $f_0$ as functions of~$E$.
This kinematic variable is related to the more common choice
$q^2=m_B^2+m_\pi^2-2m_BE$.
The contribution of $f_0$ to the decay rate is suppressed by a
factor $(m_l/m_B)^2$ so we shall present the rate given in
Eq.~(\ref{eq:dgamma}).
In the decay $B\to\pi\tau\nu$ both form factors are important, however,
so both are tabulated below, in Sec.~\ref{sec:results}.

The first determinations of $|V_{ub}|$ came from the rate of the
inclusive semileptonic decay $B\to X_ul\nu$.
In general, inclusive rates can be described model-independently 
through an operator product expansion (OPE), leading to a double 
series in $\Lambda_{\text{QCD}}/m_b$ and 
$\alpha_s(m_b)$~\cite{Ligeti:1999yc}.
Thus, they are subject to non-perturbative and perturbative
uncertainties.
In particular, one requires the quantities $\bar{\Lambda}$,
$\lambda_1$, and $\lambda_2$, which are defined in the heavy-quark
effective theory.%
\footnote{A new method for calculating $\bar{\Lambda}$, $\lambda_1$,
and $\lambda_2$ can be found in Ref.~\cite{Kronfeld:2000gk}.}
The huge charm background in $B\to X_ul\nu$ must be
eliminated by imposing a cut either on the charged lepton
energy~\cite{Albrecht:1991wa}, on the hadronic invariant
mass~\cite{Falk:1997gj}, or on $q^2$~\cite{Bauer:2000xf}.
Such cuts narrow the kinematic acceptance and may, therefore, increase
sensitivity to violations of quark-hadron duality, which is hard
to quantify.

The differential rates of exclusive decays offer an alternative route
to~$|V_{ub}|$ and~$|V_{cd}|$. 
This method is limited, however, by uncertainties in the form factors,
such as~$f_+(E)$ in Eq.~(\ref{eq:dgamma}).
In the case of $D$ decays, the $E$ dependence of the rate has
been measured only for $D\to Kl\nu$~\cite{Frabetti:1995xq}.
The FOCUS collaboration~\cite{O'Reilly:1999mp} will improve that
measurement and also should be able to measure the $E$ dependence in
the Cabibbo-suppressed mode $D\to\pi l\nu$.
First measurements of the branching ratios for $B\to\pi l\nu$
and $B\to\rho l\nu$ have been presented by the CLEO
collaboration~\cite{Alexander:1996qu}.
The form factors for all these processes are calculable with lattice
QCD.
Here we concentrate on calculating the form factors for
$B\to\pi l\nu$ (and similar $D$ decays).
The branching ratio is not as large as for $B\to\rho l\nu$, and there
are other experimental difficulties~\cite{Behrens:2000vv}.
On the other hand, with vector mesons several form factors enter into
the decay rate.
Furthermore, one might expect greater uncertainties for the $\rho$
(and $\omega$ and $\phi$) from the quenched approximation, because
of their non-zero hadronic widths.

With lattice QCD a very pressing issue is to understand the systematic
uncertainties.
Indeed, an important justification for using the quenched approximation
is that the savings in computer time allow us to study the other
systematic uncertainties in detail.
To control systematic errors we apply three main methodological
improvements in this paper: we normalize the heavy-quark action and
current in a way that reduces heavy-quark discretization effects,
we have three different lattice spacings to study any remaining
discretization effects, and we introduce kinematic cuts to avoid
model dependence.

First, let us consider the discretization for the heavy quark.
At the lattice spacings,~$a$, currently in use, the large mass of the
$b$~quark means that $m_ba>1$.
To control lattice spacing effects, we adopt the approach of
Ref.~\cite{El-Khadra:1997mp}, which takes an improved action for
Wilson fermions, but adjusts the couplings in the action and the
normalization of the current so that the leading and next-to-leading
terms in the heavy-quark effective theory (HQET) are correct.
By applying HQET directly to lattice observables, one can
show that the heavy-light meson has small discretization
effects~\cite{Kronfeld:2000ck}, in our case of order
$\alpha_s\Lambda_{\text{QCD}}/m_Q$, $\alpha_s\Lambda_{\text{QCD}}a$, 
$(\Lambda_{\text{QCD}}/m_Q)^2$, and $(\Lambda_{\text{QCD}}a)^2$.
These normalization conditions allow us to perform our
calculations directly at the physical mass $m_Q=m_b$.
This approach has already been successfully applied in
calculations of $B$ and $D$ meson decay constants by four
groups~\cite{Aoki:1998ji,El-Khadra:1998hq,Bernard:1998xi,AliKhan:2000eg}
and in calculations of the form factors for $B\to D^{(*)}l\nu$
at zero recoil~\cite{Hashimoto:2000yp}.
Work on $B\to\pi l\nu$ by two other
groups~\cite{Bowler:1995zr,Bowler:2000tx,Abada:2000ty} with the
same action (but different lattice currents) has used normalization
conditions designed for light quarks, which suffer from errors of
order $\alpha_s m_Qa$~\cite{Bowler:1995zr}
or~$(m_Qa)^2$~\cite{Bowler:2000tx,Abada:2000ty}.
To reduce these effects their calculations have been carried out with 
pseudoscalar meson masses
$1.2~\text{GeV}<m_P<2.0~\text{GeV}$~\cite{Bowler:2000tx} or
$1.7~\text{GeV}<m_P<2.6~\text{GeV}$~\cite{Abada:2000ty}.
We~have not been persuaded that HQET can be used to guide the 
extrapolation from there back up to $m_B=5.3$~GeV.

Second, there are cutoff effects of order $\alpha_ska$ and
$(ka)^2$ from the light quark, where~$k$ is the momentum of
the light quarks inside the mesons.
For light or heavy-light hadrons at rest, the momentum
$k\approx\Lambda_{\text{QCD}}$, so these effects are of the same kind
as some of those considered above.
In the semileptonic decay, however, one has a light daughter hadron
with non-zero recoil momentum, which gives rise to lattice spacing
errors with $k=|\bbox{p}_{\pi}|$.
To study this systematic error, we carry out the calculation at three 
different lattice spacings, and check the dependence of our results 
on~$a$.
We can then restrict our final results to small enough recoil momenta, 
so that discretization effects remain under control.
Our test of the lattice spacing dependence is the first in a lattice 
calculation of semileptonic form factors.

\begin{figure}[btp]
	\centering
	\epsfysize=0.45\textheight 
		\epsfbox{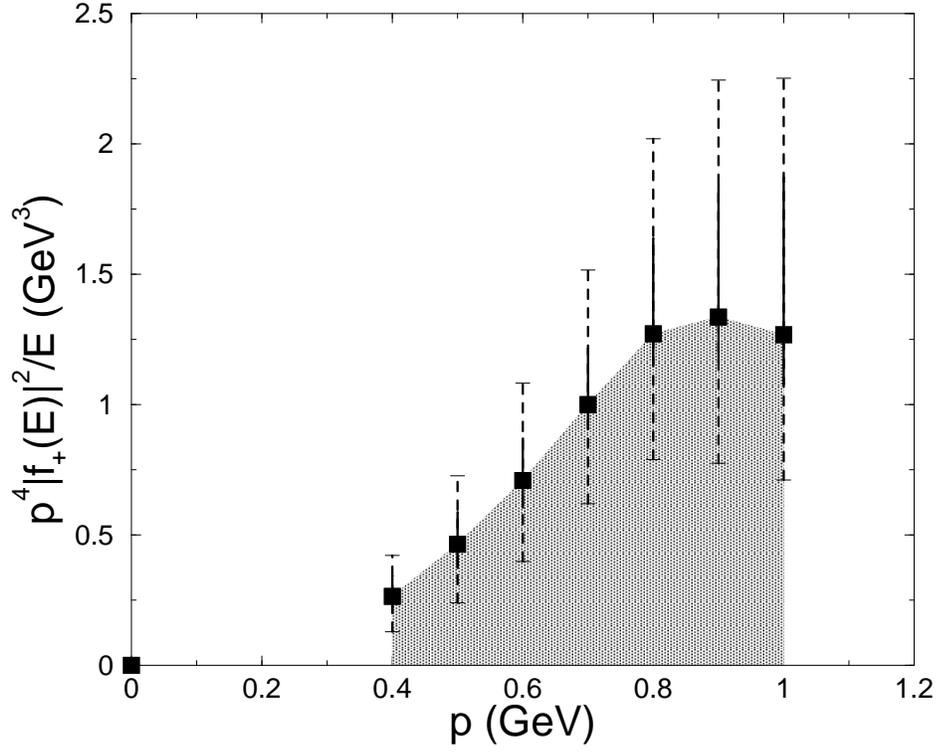}
	\epsfysize=0.45\textheight 
		\epsfbox{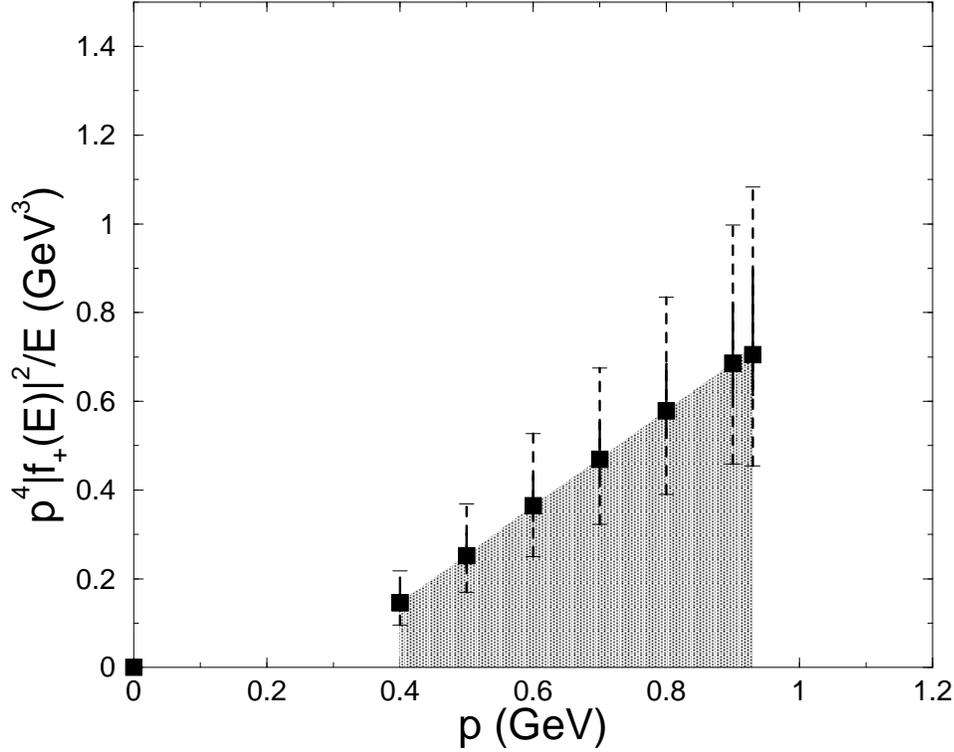}
	\caption[fig:dGammaD_b5.9]{The differential decay rate
		(without momentum-independent factors) as a function
		of~$p=|\bbox{p}_\pi|$, for (a) $B\to\pi l\nu$ and
		(b) $D\to\pi l\nu$.
		The solid error bars show the statistical uncertainty and the
		dotted ones show the sum in quadrature of statistical and
		systematic uncertainties.}
	\label{fig:dGammaD_b5.9}
\end{figure}

Third, we do not use models to extend our kinematic reach to high 
pion energy (\emph{i.e.}, low~$q^2$), in contrast to previous 
work~\cite{Bowler:1995zr,Bowler:2000tx,Abada:2000ty}.
The extrapolation would rely on the worst of our data: not only do 
discretization errors increase with $\bbox{p}_\pi a$, but statistical 
errors do too.
Therefore, we quote the differential decay rate over the range where
systematic uncertainties from the lattice are under control.
In particular, we define
\begin{equation}
	T_B(p_{\text{min}},p_{\text{max}}) =
		\int_{p_{\text{min}}}^{p_{\text{max}}} \!\!dp\;
		p^4 |f_+(E)|^2/E.
	\label{T}
\end{equation}
The upper limit is chosen to rein in the discretization and statistical
uncertainties.
The lower limit cuts out a region where extrapolations in~$p$ and
light quark mass are difficult.
Then, assuming a massless charged lepton, one can combine $T_B$ with
experimental measurements to determine
the CKM matrix via
\begin{equation}
	|V_{ub}|^2 = \frac{12\pi^3}{G_F^2m_B}
		\frac{1}{T_B(p_{\text{min}},p_{\text{max}})}
		\int_{p_{\text{min}}}^{p_{\text{max}}} \!\!dp\,
		\frac{d\Gamma_{B\to\pi}}{dp} ,
	\label{Vub}
\end{equation}
and, similarly,
\begin{equation}
	|V_{cd}|^2 = \frac{12\pi^3}{G_F^2m_D}
		\frac{1}{T_D(p_{\text{min}},p_{\text{max}})}
		\int_{p_{\text{min}}}^{p_{\text{max}}} \!\!dp\,
		\frac{d\Gamma_{D\to\pi}}{dp} .
	\label{Vcd}
\end{equation}
Our final result, showing the integrand of Eq.~(\ref{T}) for $B$~and
$D\to\pi l\nu$, is in Fig.~\ref{fig:dGammaD_b5.9}.
The shaded regions indicate the range of pion momentum over which
we can control the uncertainties.
Integrating over this region, we find
\begin{eqnarray}
	T_B(0.4~\text{GeV},1.0~\text{GeV}) & = & 0.55^{+\,0.15}_{-\,0.05}
		{\;}^{+\,0.09}_{-\,0.12}	
		{\,}^{+\,0.09}_{-\,0.02}	
		\pm 0.06					
		\pm 0.09~\text{GeV}^4,		
	\label{TB10} \\
	T_D(0.4~\text{GeV},0.925~\text{GeV}) & = & 0.23^{+\,0.04}_{-\,0.02} 
		{\;}^{+\,0.01}_{-\,0.05}	
		{\,}^{+\,0.06}_{-\,0.02}	
		\pm 0.03					
		\pm 0.03~\text{GeV}^4.		
	\label{TD09}
\end{eqnarray}
where the first uncertainty is statistical,
and following four are systematic and come from
chiral extrapolation,
lattice spacing dependence,
matching to continuum QCD,
and the sum in quadrature of several other uncertainties.
The last includes an estimate of the uncertainty from converting 
lattice units to physical units, which partly reflects uncertainty 
from the quenched approximation.
In addition to these uncertainties, which are quantifiable within the
quenched approximation, there may be an additional error from quenching
as large as 10--20~percent on~$T_B$ and~$T_D$.

At low momenta the experimental rates go to zero, so no information is 
lost by making the cut at $p_{\text{min}}=0.4$~GeV.
For semileptonic $D$ decays the high-momentum cut is already at the 
kinematic endpoint $(m_D^2-m_\pi^2)/2m_D=0.925$~GeV.
A~high-momentum cut at $p_{\text{max}}=1.0$~GeV is, however, an 
obstacle to determining~$|V_{ub}|$, since semileptonic $B$ decays 
usually produce harder pions.
Although the cut does reduce the overlap between our lattice
calculation and experimental results, the results presented here are
model independent (apart from quenching).
As experimental and lattice results improve over the next several years,
the range of pion momentum should widen and can be selected to optimize
the combined experimental and theoretical uncertainty.

This paper is organized as follows.
Section~\ref{sec:bckgnd} contains a discussion of the lattice action
and vector current for heavy quarks.
The lattice calculation of the matrix elements is described in
Sec.~\ref{sec:numerics}.
Section~\ref{sec:extend} describes an interpolation in pion
three-momentum and an extrapolation in light quark mass, which are
needed to obtain the form factors.
The former is a special feature of these decays; it interacts
with the chiral limit, and together these lead to the cuts given
in Eqs.~(\ref{TB10}) and~(\ref{TD09}).
We discuss quantitatively the systematic errors on~$T_B$ and~$T_D$
in Sec.~\ref{sec:syserr}.
The analysis of $B$ and $D$ decays is essentially the same.
Results for the form factors are tabulated in Sec.~\ref{sec:results}.
Section~\ref{sec:compare} compares our methods and results to previous
(and ongoing) work~\cite{Bowler:2000tx,Abada:2000ty,Aoki:2000ij}.
Section~\ref{sec:conclude} concludes.

Preliminary results of this analysis have been presented 
in Refs.~\cite{Simone:1996sm,Ryan:1999tj}.
Phenomenological implications of $D$ decays, especially for comparing
$D\to\pi l\nu$ and $D\to Kl\nu$ as in Ref.~\cite{Simone:1999ti}, will
appear in another publication.

\section{Continuum and lattice matrix elements}
\label{sec:bckgnd}

The continuum matrix element of the flavor-changing vector current,
${\cal V}^\mu=\bar{u}i\gamma^\mu b$, is parametrized by two independent
form factors, for example those in Eq.~(\ref{eq:ff}).
In considering the chiral and heavy-quark limits, it is more convenient
to write the matrix element as
\begin{equation}
	\langle\pi (p_\pi)|{\cal V}^\mu |B(p_B)\rangle =
		\sqrt{2m_B} \left[v^\mu f_\parallel(E) + 
		p^\mu_\perp f_\perp(E) \right],
	\label{ff_vp} 
\end{equation}
where $v=p_B/m_B$ is the four-velocity of the $B$, and
$p_\perp=p_\pi-Ev$ is the pion momentum orthogonal to~$v$.
The traditional form factors~$f_+$ and~$f_0$ are related
to~$f_\parallel$ and~$f_\perp$ by
\begin{eqnarray}
    f_+(E) & = & (2m_B)^{-1/2}
		\left[f_\parallel(E) + (m_B-E)f_\perp(E) \right],
	\label{f+} \\
    f_0(E) & = & \frac{\sqrt{2m_B}}{m_B^2-m_\pi^2}
		\left[ (m_B-E)f_\parallel(E) + (E^2-m_\pi^2)f_\perp(E) \right].
	\label{f0}
\end{eqnarray}
At $q^2=0$ it follows from these formulae that
$f_+=f_0$, which is necessary from Eq.~(\ref{eq:ff}).

There are several good reasons to focus the numerical analysis
on~$f_\parallel$ and~$f_\perp$.
First, consideration of chiral and heavy-quark symmetry yields the
expectation for $m_\pi$, $E\to 0$
\begin{eqnarray}
    f_\parallel & = & \frac{f_B\sqrt{m_B}}{\sqrt{2}f_\pi},
	\label{soft_parallel} \\
    f_\perp     & = & \frac{f_{B^*}\sqrt{m_{B^*}}}{\sqrt{2}f_\pi}
		g_{BB^*\pi} \frac{2m_B}{m_{B^*}^2-q^2},
	\label{soft_perp}
\end{eqnarray}
through order $1/m_Q$ in the heavy-quark
expansion~\cite{Burdman:1994es}.
Here $f_B$, $f_{B^*}$, and $f_\pi$ are decay constants, 
and $g_{BB^*\pi}$ is the $B$-$B^*$-$\pi$ coupling.
Although we do not use these results to constrain the needed chiral
extrapolation of our data, they do show us that $f_\parallel$ and
$f_\perp$ behave differently as $m_\pi$ is reduced to its physical
value.
(Recall $q^2=m_B^2+m_\pi^2-2m_BE$.)
Furthermore, $f_\parallel$ and $f_\perp$ have a simple description
in the heavy-quark effective theory~\cite{Burdman:1994es}, so
they are natural quantities to study in the lattice method of
Refs.~\cite{El-Khadra:1997mp,Kronfeld:2000ck},
or when using lattice NRQCD~\cite{Aoki:2000ij}.
Finally, they emerge directly from the lattice calculation, so it is
simpler to analyze them separately, forming the linear combinations
in Eqs.~(\ref{f+}) and~(\ref{f0}) at the end.

For the light quarks we use the Sheikholeslami-Wohlert (SW)
action~\cite{Sheikholeslami:1985ij}, with the customary normalization
conditions for $m_qa\to 0$.
The SW action has an extra coupling $c_{\text{SW}}$, sometimes called 
the ``clover'' coupling, which can be adjusted to reduce the leading 
lattice-spacing effect of Wilson fermions.
In practice, we adjust~$c_{\text{SW}}$ according to tadpole-improved,
tree-level perturbation theory~\cite{Lepage:1993xa}, so the leading 
light-quark cutoff effect is of order $\alpha_s ka$.

We also use the SW action for the heavy quark, but its two free 
parameters, the bare mass~$m_0$ and clover coupling~$c_{\text{SW}}$, 
are adjusted to maintain good behavior in the heavy-quark 
limit~\cite{El-Khadra:1997mp}.
This goes as follows:
on-shell lattice matrix elements can be described by a version of 
HQET~\cite{Kronfeld:2000ck}, with effective Lagrangian
(in the rest frame)
\begin{equation}
	{\cal L}_{\text{HQET}} = m_1\bar{h}h +
		\frac{\bar{h}\bbox{D}^2h}{2m_2} +
		\frac{\bar{h}\,i\bbox{\Sigma}\cdot\bbox{B}\,h}{2m_{\cal B}}
		+ \cdots,
\end{equation}
where $h$ is the heavy-quark field of~HQET, and $\bbox{B}$ is the
chromomagnetic field.
The ``masses'' $m_1$, $m_2$, and $m_{\cal B}$ are short-distance
coefficients; they depend on $m_0$ and $c_{\text{SW}}$ (and the gauge
coupling).
Fortunately, matrix elements are completely independent
of~$m_1$~\cite{Kronfeld:2000ck}, so we adjust $m_0$ and $c_{\text{SW}}$
to tune $m_2$ and $m_{\cal B}$ to the $b$ (or $c$) quark.
In practice, we tune $m_2$ non-perturbatively, using the quarkonium
spectra, and $m_{\cal B}$ with the estimate of tadpole-improved,
tree-level perturbation theory~\cite{Lepage:1993xa}.

The lattice current is constructed according to the same principles.
We distinguish the lattice current $V$ from its continuum
counterpart~${\cal V}$ and take
\begin{equation}
	V^\mu = \sqrt{Z_{V^{uu}} Z_{V^{bb}}}
		\bar{\Psi}_ui\gamma^\mu \Psi_b
	\label{Vlat}
\end{equation}
where the rotated field~\cite{El-Khadra:1997mp}
\begin{equation}
	\Psi_q = \left[1 + ad_1
	\bbox{\gamma}\cdot\bbox{D}_{\text{lat}}\right] \psi_q,
	\label{rotate}
\end{equation}
and $\psi_q$ is the lattice quark field ($q=u,\;b$) in the SW action.
Here $\bbox{D}_{\text{lat}}$ is the symmetric, nearest-neighbor, 
covariant difference operator.
In Eq.~(\ref{Vlat}) the factors $Z_{V^{qq}}$, $q=u,b$, normalize the
flavor-conserving currents.
In practice, they are computed non-perturbatively.

Matching the current~$V^\mu$ to HQET requires further short-distance
coefficients:
\begin{equation}
	V^\mu \doteq
		(\eta_V^{\text{lat}}+\zeta_V^{\text{lat}}) v^\mu \bar{q}h  +
		\eta_V^{\text{lat}} \bar{q}i\gamma^\mu_\perp h
	 -	\frac{\bar{q}i\gamma^\mu{\kern+0.1em /\kern-0.65em D}_\perp h}{2m_3}
		+ \cdots,
	\label{VlatHQET}
\end{equation}
where the symbol $\doteq$ implies equality of matrix elements, and
$\bar{q}$ is a relativistic (continuum) anti-quark field.
At the tree level $\eta_V^{\text{lat}}=1$, $\zeta_V^{\text{lat}}=0$.
Also, further dimension-four operators, whose coefficients
vanish at the tree level, are omitted from the right-hand side of
Eq.~(\ref{VlatHQET}).
This description is in complete analogy with that for the continuum 
current, namely,
\begin{equation}
	{\cal V}^\mu \doteq (\eta_V+\zeta_V) v^\mu \bar{q}h  +
		\eta_V \bar{q}i\gamma^\mu_\perp h
	 -	\frac{B_1\bar{q}i\gamma^\mu{\kern+0.1em /\kern-0.65em D}_\perp h}{2m_Q}
		+ \cdots.
	\label{VcontHQET}
\end{equation}
Indeed, the HQET operators are the same.
On the other hand, the radiative corrections to the short-distance
coefficients in Eqs.~(\ref{VlatHQET}) and Eqs.~(\ref{VcontHQET})
differ, because the lattice modifies the physics at short distances.

By studying the form factors in HQET, as in Ref.~\cite{Burdman:1994es},
one can deduce how to compensate for the mismatch between short-distance
coefficients $\eta_V^{\text{(lat)}}$ and $\zeta_V^{\text{(lat)}}$ for
the lattice and $\eta_V$ and $\zeta_V$ for the continuum.
HQET matrix elements have form factors
\begin{eqnarray}
	\langle\pi|\bar{q}h|B\rangle & = &
		\varphi_\parallel(E) , \\
	\langle\pi|\bar{q}i\gamma_\perp^\mu h|B\rangle & = &
		p_\perp^\mu \varphi_\perp(E),
\end{eqnarray}
so, leaving aside the dimension-four operator
$\bar{q}i\gamma^\mu{\kern+0.1em /\kern-0.65em D}_\perp h$ for now,
\begin{eqnarray}
		f_\parallel(E) & = & \eta_V \varphi_\parallel(E) ,
		\label{para} \\
		f_\perp(E) & = & (\eta_V+\zeta_V) \varphi_\perp(E) .
		\label{perp}
\end{eqnarray}
By the same reasoning, form factors calculated with the lattice
current~$V$ satisfy
\begin{eqnarray}
		f^{\text{lat}}_\parallel(E) & = &
			\eta^{\text{lat}}_V \varphi_\parallel(E) , 
		\label{paralat} \\
		f^{\text{lat}}_\perp(E) & = &
			(\eta^{\text{lat}}_V+\zeta^{\text{lat}}_V) \varphi_\perp(E) .
		\label{perplat}
\end{eqnarray}
Up to lattice artifacts of the light degrees of freedom the HQET form 
factors~$\varphi_\parallel$ and $\varphi_\perp$ are the same 
in Eqs.~(\ref{para}) and~(\ref{perp}) and
in Eqs.~(\ref{paralat}) and~(\ref{perplat}).
Thus,
\begin{eqnarray}
		f_\parallel(E) & = & \rho_{V_\parallel} 
			f^{\text{lat}}_\parallel(E), \label{f=rfparl} \\
		f_\perp(E)     & = & \rho_{V_\perp} 
			f^{\text{lat}}_\perp(E),     \label{f=rfperp} 
\end{eqnarray}
where $\rho_{V_\parallel}=\eta_V/\eta^{\text{lat}}_V$,
$\rho_{V_\perp}=(\eta_V+\zeta_V)/%
(\eta^{\text{lat}}_V+\zeta^{\text{lat}}_V)$.
Because these factors arise from short distances, in practice we
compute them in perturbation theory to one loop.
We find these short-distance corrections to be very small.

Finally, the free parameter $d_1$ in Eq.~(\ref{rotate}) can be adjusted
to tune $1/m_3$ to $B_1/m_Q$.
In the present calculations, we adjust~$d_1$ with the estimate of
tadpole-improved, tree-level perturbation theory, as explained in 
Ref.~\cite{El-Khadra:1997mp}.

With these normalization conditions the leading term in the heavy-quark
expansion is correctly obtained, up to neglected higher-order
corrections to $\rho_{V_\parallel}$ and~$\rho_{V_\perp}$.
The associated error should be much smaller than our
other uncertainties, because most of the short-distance
normalization is handled non-perturbatively, through the factor
$\sqrt{Z_{V^{uu}}Z_{V^{bb}}}$.
Similarly, the $1/m_Q$ term in the heavy-quark expansion is correctly 
obtained, up to neglected loop corrections to $c_{\text{SW}}$ and 
$d_1$, and to dimension-four operators neglected in 
Eq.~(\ref{VlatHQET}).
Here the associated error depends on~$m_Qa$.
When $m_Qa>1$ it is formally of order
$\alpha_s\Lambda_{\text{QCD}}/m_Q$, but when $m_Qa<1$ it is formally
of order $\alpha_s\Lambda_{\text{QCD}}a$.
In the work reported here, such corrections are smaller than,
or comparable to, other uncertainties.

In lattice QCD the required matrix elements and thence the form
factors are calculated from correlation functions.
In particular, the three-point correlation function
for the $B\to\pi$ transition is
\begin{equation}
	C_\mu^{(3)}(\bbox{p},\bbox{k}, t_f,t_s,t_i) =
	\sum_{\bbox{x},\bbox{y}}
	e^{-i\bbox{p}\cdot\bbox{x}}
	e^{-i(\bbox{k}-\bbox{p})\cdot\bbox{y}}
	\langle 0| {\cal O}_B(\bbox{x},t_f)
	\bar{\Psi}_b\gamma_\mu\Psi_u(\bbox{y},t_s)
	{\cal O}^\dagger_\pi(\bbox{0}, t_i)|0\rangle
	\label{eq:3ptfunc},
\end{equation}
where ${\cal O}_B$ and ${\cal O}_\pi$ are interpolating operators
for the $B$ and $\pi$ mesons.
In the limit of large time separations, the correlation function becomes 
\begin{equation}
	C_\mu^{(3)}(\bbox{p},\bbox{k}, t_f,t_s,t_i) = 
		{\cal Z}_B^{1/2} {\cal Z}_\pi^{1/2}
	\frac{\langle B(\bbox{k})|
	\bar{\Psi}_b\gamma_\mu\Psi_u
	|\pi(\bbox{p})\rangle}%
	{\sqrt{2E_B}\hfil\sqrt{2E_\pi}}
		e^{-E_\pi(t_s-t_i)}e^{-E_B(t_f-t_s)} + \cdots,
	\label{three-pt}
\end{equation}
where $E_B$ ($E_\pi$) is the energy of a $B$ ($\pi$) meson with momentum
$\bbox{k}$ ($\bbox{p}$).
The energies and the external line factors ${\cal Z}_\pi$ and
${\cal Z}_B$ can be calculated from two-point correlation functions
\begin{equation}
	C^{(2)}(\bbox{p}, t) = \sum_{\bbox{x}}
		e^{-i\bbox{p}\cdot\bbox{x}}
		\langle 0|{\cal O}_H(\bbox{x},t)
		{\cal O}^\dagger_H(\bbox{0},0)|0\rangle,
\end{equation}
where $H$ is $\pi$ or $B$, and for large $|t|$ one has
\begin{equation}
	C^{(2)}(\bbox{p},t) = {\cal Z}_H e^{-E_H|t|} + \cdots.
	\label{two-pt}
\end{equation}
By time reversal
$\langle B(\bbox{k})|\bar{\Psi}_b\gamma_\mu\Psi_u|\pi(\bbox{p})\rangle=%
\langle\pi(\bbox{p})|\bar{\Psi}_u\gamma_\mu\Psi_b|B(\bbox{k})\rangle$, so 
in the rest of this paper we do not distinguish the two matrix elements.

To summarize this section, let us review the steps needed to obtain the
physical form factors~$f_+$ and $f_0$.
First we obtain $E_\pi$ and
\begin{eqnarray}
	F_\parallel(\bbox{p}) & = & \frac{\langle B(\bbox{0})|%
		\bar{\Psi}_b\gamma_4\Psi_u|\pi(\bbox{p})\rangle}%
		{\sqrt{2m_B}\hfil\sqrt{2E_\pi}},
	\label{eq:Fpara} \\
	F_\perp(\bbox{p}) & = & \frac{1}{p_ja} \frac{\langle B(\bbox{0})|%
		\bar{\Psi}_b\gamma_j\Psi_u|\pi(\bbox{p})\rangle}%
		{\sqrt{2m_B}\hfil\sqrt{2E_\pi}},
	\label{eq:Fperp}
\end{eqnarray}
for several values of~$\bbox{p}$, directly from fitting the lattice
correlation functions to the time dependence given in
Eqs.~(\ref{three-pt}) and~(\ref{two-pt}).
The normalization factors $Z_{V^{uu}}$ and $Z_{V^{bb}}$ are computed
from zero-momentum, flavor-conserving correlation functions.
The radiative correction factors $\rho_V$ appearing in
Eqs.~(\ref{f=rfparl}) and~(\ref{f=rfperp}) are computed with 
perturbation theory.
These ingredients are combined to form
\begin{eqnarray}
	f_\parallel(E) & = & \rho_{V_\parallel}
		\sqrt{Z_{V^{uu}} Z_{V^{bb}}}\sqrt{2E_\pi}\,
		F_\parallel(\bbox{p}) , \\
	f_\perp(E)     & = & \rho_{V_\perp}
		\sqrt{Z_{V^{uu}} Z_{V^{bb}}}\sqrt{2E_\pi}\, a
		F_\perp(\bbox{p}) ,
\end{eqnarray}
with $E=E_\pi$.
From the calculated values of $\bbox{p}$ we then interpolate to a
fiducial set of momenta.
The form factors $f_\parallel$ and $f_\perp$ are extrapolated to the
physical light quark mass.
With the light quark corresponding to strange we check also for 
lattice spacing effects.
Finally, the combinations $f_+$ and $f_0$ are formed
from the extrapolated $f_\perp$ and $f_\parallel$ with
Eqs.~(\ref{f+}) and~(\ref{f0}) and physical meson masses.

\section{Lattice calculation}
\label{sec:numerics}

This work uses three ensembles of lattice gauge field configurations,
which have been used in previous work on heavy-light decay
constants~\cite{Duncan:1995uq,El-Khadra:1998hq}, light-quark
masses~\cite{Gough:1997kw}, and quarkonia~\cite{El-Khadra:1992vn}.
The quark propagators are the same as in Ref.~\cite{El-Khadra:1998hq}, 
but we now use 200 instead of 100 configurations on the finest lattice 
(with $\beta=6.1$).
The input parameters for these fields are in
Table~\ref{tab:sim_details}, together with some elementary output
parameters.
\begin{table}
\centering
\caption[tab:sim_details]{Input parameters to the numerical lattice
calculations, together with some elementary output parameters.
Error bars on the outputs refer to the last digit(s).}
\label{tab:sim_details} 
\begin{tabular}{lccc}
\multicolumn{4}{c}{Inputs} \\
\hline
$\beta=6/g_0^2$ &6.1 &5.9 &5.7 \\
Volume, $N_S^3\times N_T$ &$24^3\times48$ &$16^3\times32$ &$12^3\times24$\\
Configurations                  &200  &350 &300\\ 
$c_{\rm sw}$                    &1.46 &1.50 &1.57\\ 
$\kappa_b$, $m_0$~(GeV) &0.0990, 4.31  &0.0930, 3.73  &0.0890, 2.87  \\ 
$\kappa_c$, $m_0$~(GeV) &0.1260, 1.07  &0.1227, 1.05  &0.1190, 0.96  \\
$\kappa_s$, $m_0$~(GeV) &0.1373, 0.092 &0.1385, 0.091 &0.1405, 0.093 \\ 
$\kappa_q$, $m_0$~(GeV) &              &0.1382, 0.107 & \\
                        &              &0.1388, 0.075 &0.1410, 0.076 \\ 
                        &              &0.1391, 0.059 &0.1415, 0.059 \\ 
                        &              &0.1394, 0.043 &0.1419, 0.045 \\ 
\hline
\multicolumn{4}{c}{Elementary outputs} \\
\hline
$\kappa_{\text{crit}}$        & 0.13847\er{4}{2} & 0.14021\er{3}{1} &
	0.14327\er{5}{2} \\
$a^{-1}_{\text{1P-1S}}$ (GeV) & 2.64\err{17}{13} & 1.80\er{7}{6}    &
	1.16\er{3}{3}    \\ 
$a^{-1}_{f_\pi}$ (GeV)        & 2.40\err{10}{12} & 1.47\er{6}{6}    &
	0.89\er{2}{2}    \\ 
$2\pi/N_Sa$ (GeV)             & 0.686  & 0.707  & 0.607  \\
$u_0$                         & 0.8816 & 0.8734 & 0.8608 \\
$\alpha_V(2/a)$			      & 0.171  & 0.192  & 0.227  \\
\end{tabular}
\end{table}

The quark propagators are computed from the Sheikholeslami-Wohlert
action, which includes a dimension-five interaction with
coupling~$c_{\text{SW}}$.
For heavy and light quarks we adjust~$c_{\text{SW}}$ to the value 
$u_0^{-3}$ suggested by tadpole-improved, tree-level perturbation 
theory, and the so-called mean link $u_0$ is calculated from the 
plaquette.
The hopping parameter~$\kappa$ is related to the bare quark mass.
For bottom and charmed quarks, $\kappa_b$ and $\kappa_c$ are adjusted
so that the spin-averaged kinetic mass of the corresponding 1S
quarkonium states match experimental measurements.
For light quarks, $\kappa_s$ and $\kappa_q$ are fixed from light
meson spectroscopy, using leading-order chiral perturbation theory
and the experimental kaon and pion masses.
We also list the tadpole-improved bare quark mass in GeV,
\begin{equation}
	m_0a = \frac{1}{u_0} \left(\frac{1}{2\kappa}
		- \frac{1}{2\kappa_{\text{crit}}}\right),
\end{equation}
where the critical quark hopping parameter $\kappa_{\text{crit}}$
makes the pion massless.
Although this mass is just a bare mass, it shows that the heavy quarks
are heavy, and the light quarks~light.

We calculate the three-point function in Eq.~(\ref{three-pt}) with
degenerate spectator and daughter light quarks.
At each lattice spacing we have propagators corresponding to the
strange quark.
We refer to this decay as $B_s\to\eta_s l\nu$, writing $\eta_s$ for 
the pseudoscalar $\bar{s}s$ state in analogy with quarkonium.
At $\beta = 5.9$ and $5.7$ we have additional light quark propagators,
with hopping parameter~$\kappa_q$, covering the range
$\case{1}{2}m_s\lesssim m_q\lesssim m_s$.

The lattice spacing $a$ in physical units must be set through some
fiducial observable.
As a rule~\cite{El-Khadra:1992vn} we prefer the spin-averaged 1P-1S 
splitting of charmonium,~$\Delta m_{\text{1P-1S}}$.
For comparison we give the value of $a^{-1}$ defined through the pion
decay constant~$f_\pi$.
The discrepancy means that $\Delta m_{\text{1P-1S}}/f_\pi$ does
not agree with experiment; this is thought to be largely due to the
quenched approximation, because it remains even as $a$ is decreased.

The renormalized strong coupling $\alpha_V(2/a)$ at scale $2/a$ is
determined as in Ref.~\cite{Lepage:1993xa}.
It is an ingredient in the calculation of the short-distance
coefficients $\rho_{V_\parallel}$ and $\rho_{V_\perp}$, introduced in
Eqs.~(\ref{f=rfparl}) and~(\ref{f=rfperp}).

In the three-point functions the heavy-light meson is at rest, while
the momentum of the light daughter meson is varied.
In a finite volume only discrete values of spatial momentum are
accessible.
We compute the three-point function with
$\bbox{p}_\pi=2\pi\bbox{n}/N_Sa$, for integer momentum
$\bbox{n}\in\{(0,0,0),\,(1,0,0),\,(1,1,0),\,(1,1,1),\,(2,0,0)\}$.
As one can see in Table~\ref{tab:sim_details}, one unit of momentum
is about 0.7~GeV in the boxes used here, so our calculations cover
the range $0\leq p<1.5$~GeV.

We obtain the energies, matrix elements, and ${\cal Z}_H$~factors
by fitting Eqs.~(\ref{three-pt}) and~(\ref{two-pt}) with a
$\chi^2$-minimization algorithm.
Statistical errors, including the full correlation matrix in~$\chi^2$,
are determined from 1000 bootstrap samples for each best fit.
The bootstrap procedure is repeated with the same sequence for all
quark mass combinations
and momenta, and in this way the fully correlated statistical errors
are propagated through later stages of the analysis.

\begin{figure}[btp]
	\centering
	\epsfxsize=0.45\textwidth
		\epsfbox{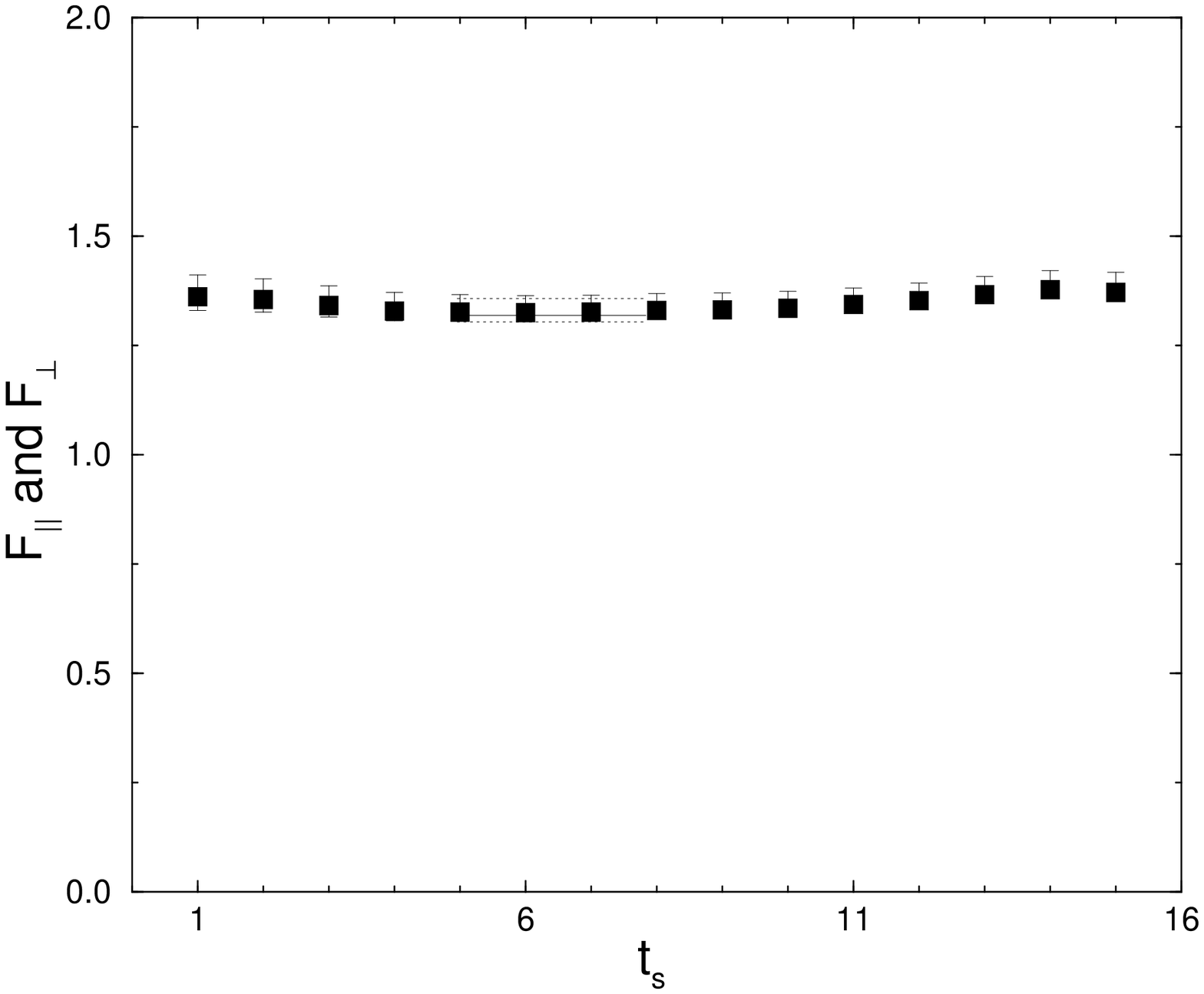}
	\epsfxsize=0.45\textwidth
		\epsfbox{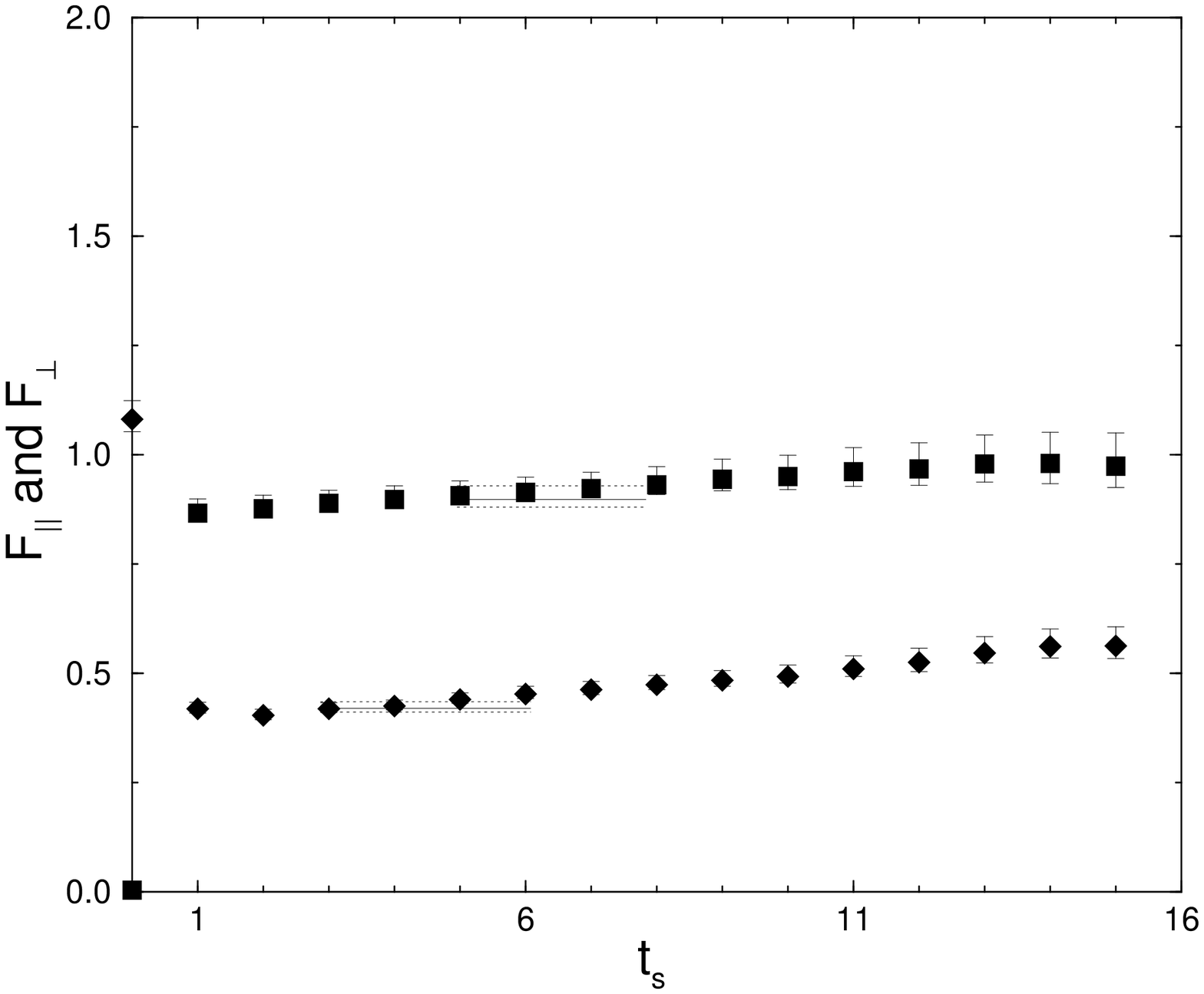}
	\epsfxsize=0.45\textwidth
		\epsfbox{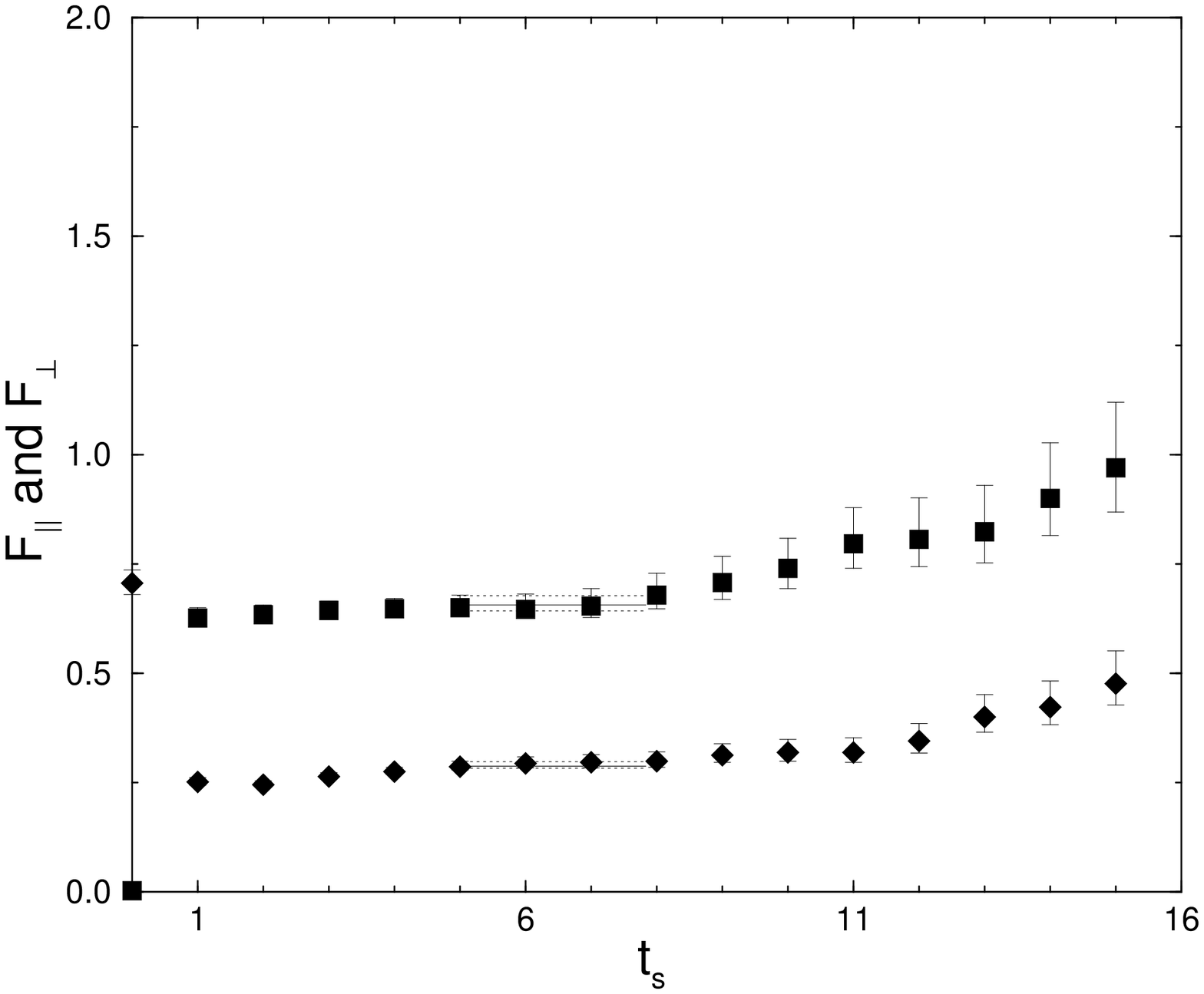}
	\epsfxsize=0.45\textwidth
		\epsfbox{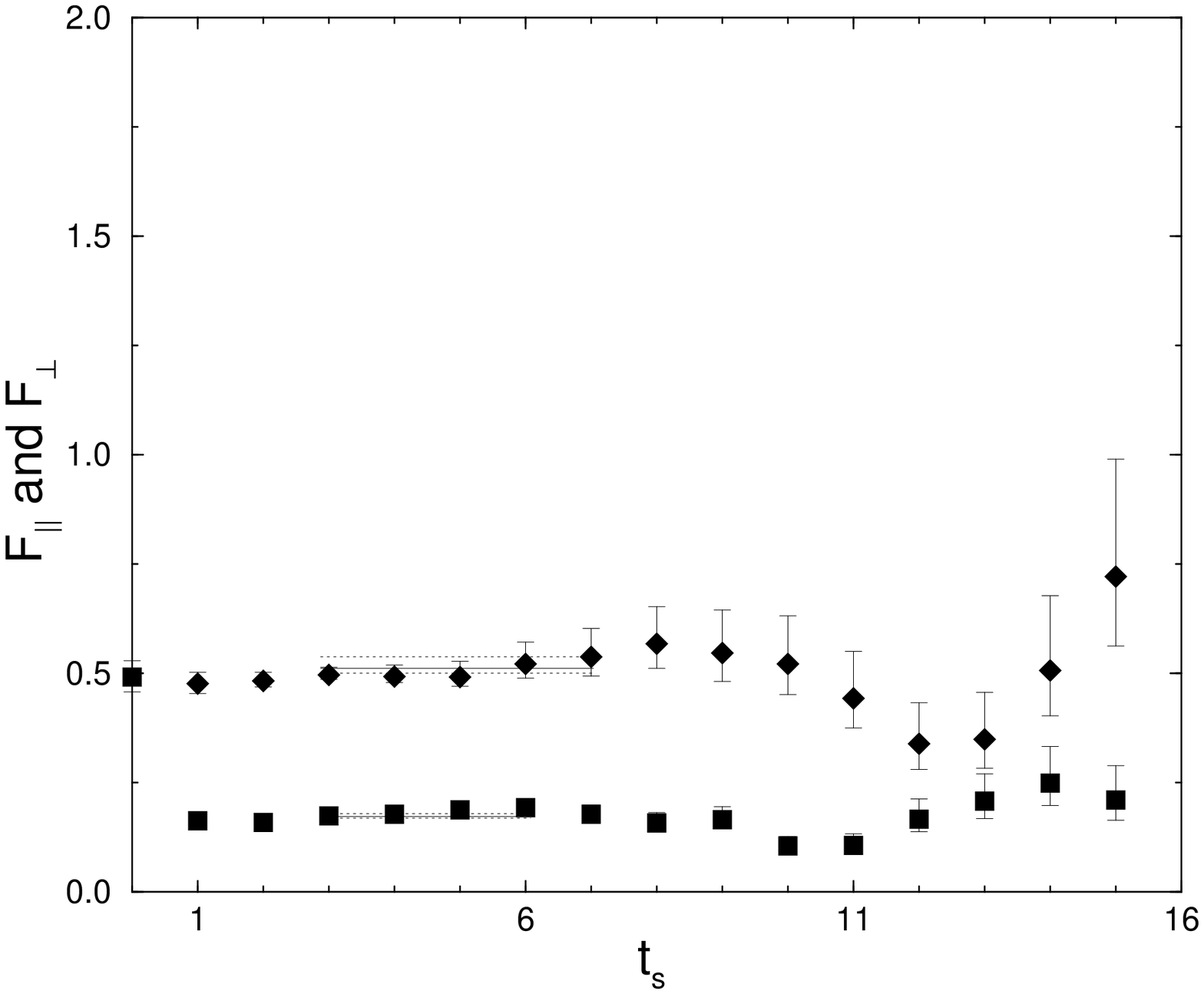}
	\caption{Matrix elements $F_\parallel$ (squares) and 
		$F_\perp$ (diamonds) at $\beta = 5.9$,
		for a light strange quark and a heavy bottom quark.
		The integer momenta are (a)~$\bbox{n}=(0,0,0)$,
		(b)~$\bbox{n}=(1,0,0)$, (c)~$\bbox{n}=(1,1,0)$, and
		(d)~$\bbox{n}=(1,1,1)$.}
	\label{fig:fits}
\end{figure}

\begin{figure}[btp]
	\centering
	\epsfxsize=0.45\textwidth
		\epsfbox{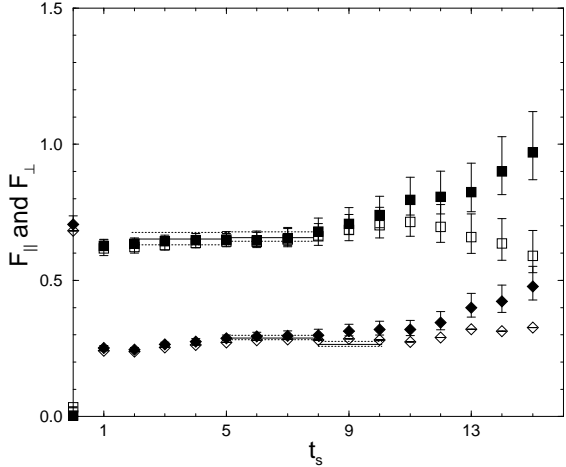}
    \caption[fig2:1Sfits]{Isolation of the lowest-lying states
		with different smearing functions, for $\bbox{n}=(1,1,0)$
		and quark masses as in Fig.~\ref{fig:fits}.
		The solid symbols have the standard 1S source and 1S sink;
		the open  symbols have a $\delta$-function sink for the~$B_s$.}
	\label{fig:1Sfits}
\end{figure}

The right-hand sides of Eqs.~(\ref{three-pt}) and~(\ref{two-pt})
are the first term in a series, with another term for each radial
excitation.
We reduce contamination from these states two ways.
First, we keep the three points of the three-point function well
separated in (Euclidean) time.
The light meson creation operator ${\cal O}_\pi$ is always at $t_i=0$
and the heavy-light meson annihilation operator at $t_f=N_T/2$.
We then vary the time $t_s$ of the current and the range $\Delta t$
of time-slices kept in the fit, to see when the lowest-lying states
dominate.
The final choice is made by demanding that $\chi^2/\text{d.o.f.}$
is acceptable and, then, minimizing the statistical errors while
maximizing~$\Delta t$.
For acceptable fits we have $3\leq\Delta t\leq 6$.
The extraction of the desired matrix elements is shown in
Fig.~\ref{fig:fits} for several light-meson momenta and typical
quark mass.
The best fit and error envelope are indicated by the solid and dotted
lines respectively.
The second way to isolate the lowest-lying states is to choose
interpolating operators, ${\cal O}_B$ and ${\cal O}_\pi$ in
Eq.~(\ref{eq:3ptfunc}), to have a large overlap with the desired
state.
This is done by smearing out the quark and anti-quark with 1S and 2S
Coulomb-gauge wave functions, as in Ref.~\cite{Duncan:1993eb}.
We also examine point-like, or $\delta$ function, operators, but for
light mesons at higher momenta we find that the $\delta$ source does
not yield good plateaus~\cite{Onogi:1995wx}.
The different combinations of sources and sinks allow
us to check explicitly for excited state contributions by comparing
results from fits with different smearing functions.
Figure~\ref{fig:1Sfits} compares results for the matrix element
$\langle B_s|V^\dagger_\mu|\eta_s\rangle$ at $\bbox{n}=(1,1,0)$,
obtained from 1S source and sink and from 1S source for the light
meson and $\delta$ sink for the~$B_s$.
The 1S-1S correlation functions yield the cleanest matrix elements, 
so we take our central values from them.

\section{Analysis of form factors}
\label{sec:extend}

From the exponential fits to three-point correlation functions described in 
Sec.~\ref{sec:numerics} we have the matrix element,
$\langle\eta_q(\bbox{p})|V^\mu|B_q(\bbox{0})\rangle$,
for quark masses $m_q\lesssim m_s$ and
final-state momenta $|\bbox{p}|<1.4$~GeV.
We must now extend these data to lower quark mass, until the mass of the
$\bar{q}q$ pseudoscalar reaches the pion mass.
Furthermore, the more important form factor $f_\perp(E)$, which is
essentially $\langle\eta_q(\bbox{p})|V_j|B_q(\bbox{0})\rangle/p_j$,
is directly calculated only for non-zero three-momentum.
In the finite volume used here, the lowest non-zero momentum is already
$0.7$~GeV, and we would like to extend to lower values, calling for
another extrapolation.

The extrapolation in quark mass can be guided by chiral perturbation
theory.
To extrapolate in momentum, however, there is no firm theoretical
guide, so we must exercise caution.
Fortunately, this extrapolation is problematic only in the kinematic 
regime where phase space suppresses the rate.
Consequently, neither extrapolation introduces a model.
We also have checked that the order in which the momentum and chiral 
extrapolations are done has no significant effect the final result.

\subsection{Momentum interpolation and extrapolation}
\label{sec:mom}
\begin{figure}[btp]
	\centering
	\epsfysize=0.45\textheight 
		\epsfbox{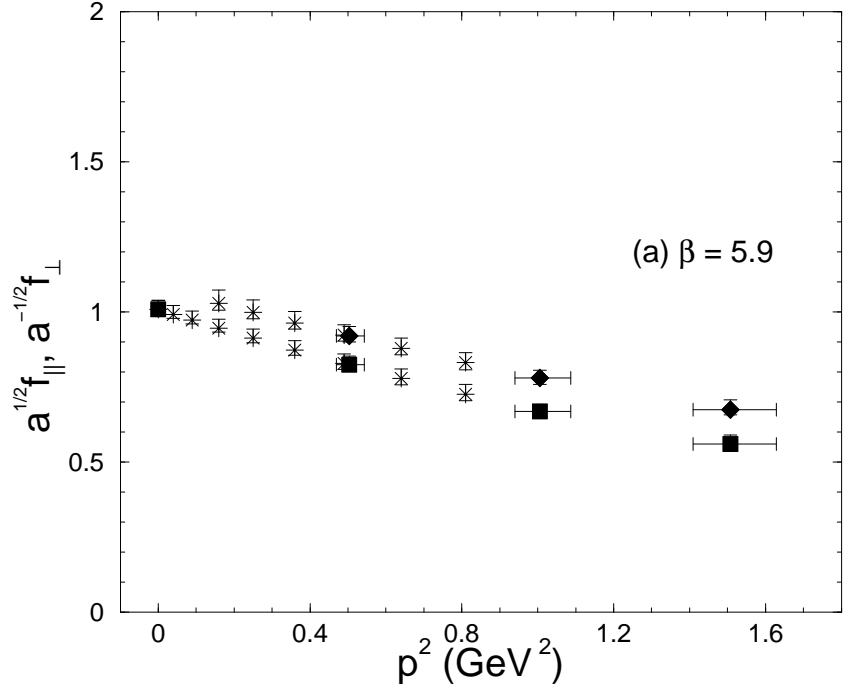}
	\epsfysize=0.45\textheight 
		\epsfbox{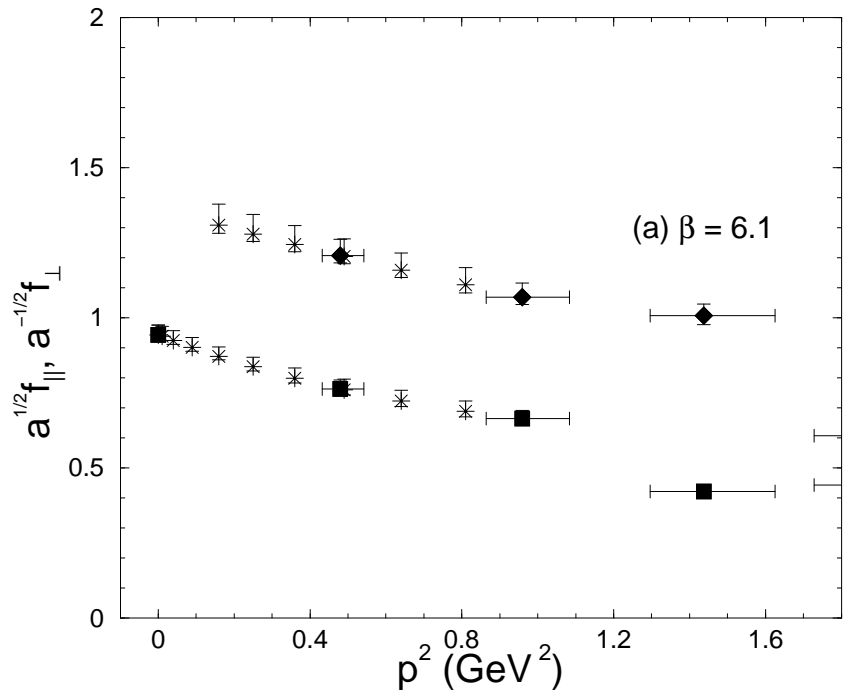}
	\caption[fig3:typ_interps6.1]{Momentum interpolation (and 
		extrapolation) of $f_\parallel a^{+1/2}$ (squares) and
		$f_\perp a^{-1/2}$ (diamonds)
		for $B_s\to\eta_s l\nu$ at 
		(a) $\beta = 5.9$, (b) $\beta = 6.1$.
		The solid points are the underlying data; the asterisks
		are interpolated.}
	\label{fig:typ_interps6.1}
\end{figure}

Ultimately, we want to compare results at the three different lattice 
spacings.
Therefore, we interpolate the lattice data to a fixed set of physical
momenta.
To start, we convert the lattice data to physical units using
$a^{-1}_{\text{1P-1S}}$.
Figure~\ref{fig:typ_interps6.1} shows the underlying data for
$B_s\to\eta_s l\nu$ at $\beta=5.9$ and $6.1$, along with interpolated 
points.
The vertical (horizontal) error bars on the underlying
data come from the statistical uncertainty in
$F_{\parallel,\perp}$~($a^{-1}$).
We interpolate 
$\log f_\perp a^{-1/2}$ ($\log f_\parallel a^{+1/2}$)
linearly (quadratically) in~$\bbox{p}^2$ to
$|\bbox{p}|\in\{0,0.1,0.2,0.3,0.4,0.5,0.6,0.7,0.8,0.9,1.0,1.1\}$~GeV.
This set forms the basis of all further analysis.
The statistical error bars of the interpolated points are vertical
only, because both statistical errors are propagated through the
interpolation.

\begin{figure}[btp]
	\centering
	\epsfysize=0.45\textheight
		\epsfbox{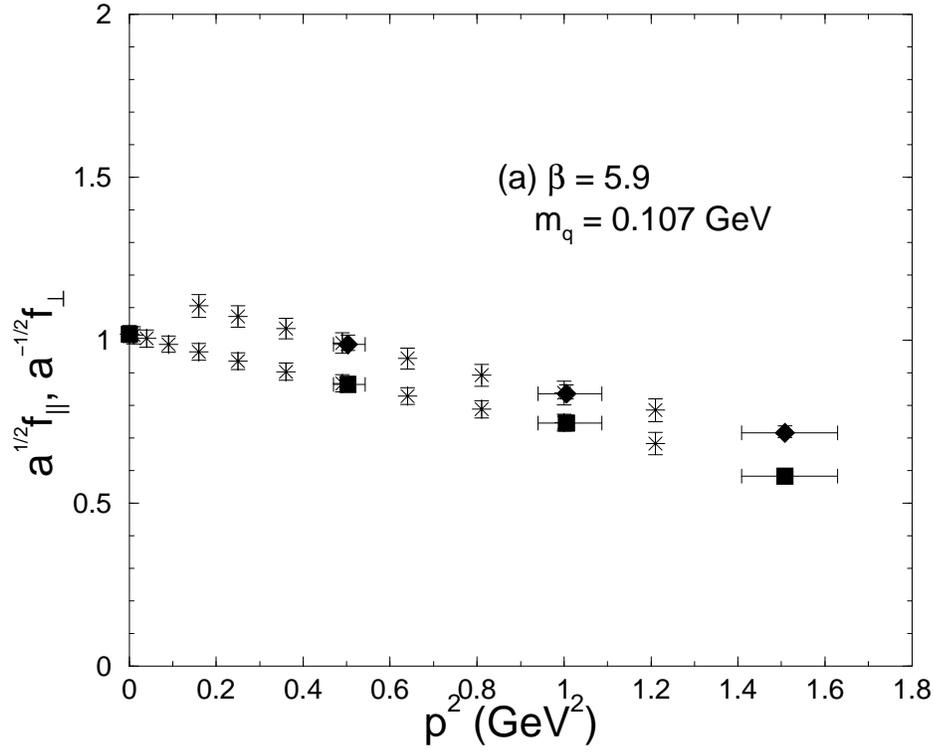}
	\epsfysize=0.45\textheight
		\epsfbox{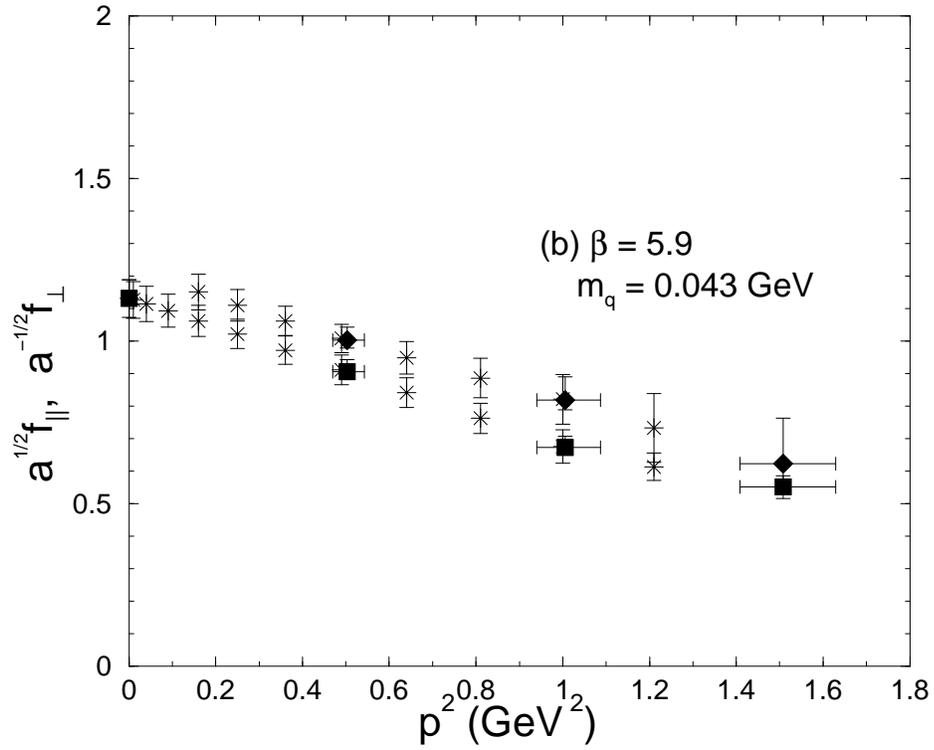}
	\caption[pole_effect]{Momentum interpolation (and extrapolation) 
		at $\beta=5.9$, for $B_q\to\eta_q l\nu$ and
		(a) the heaviest of our light quarks, with $\kappa_q = 0.1382$;
		(b) the lightest of the light quarks, with $\kappa_q = 0.1394$.}
	\label{fig:pole_effect}
\end{figure}

\begin{figure}[btp]
	\centering
	\epsfysize=0.45\textheight
		\epsfbox{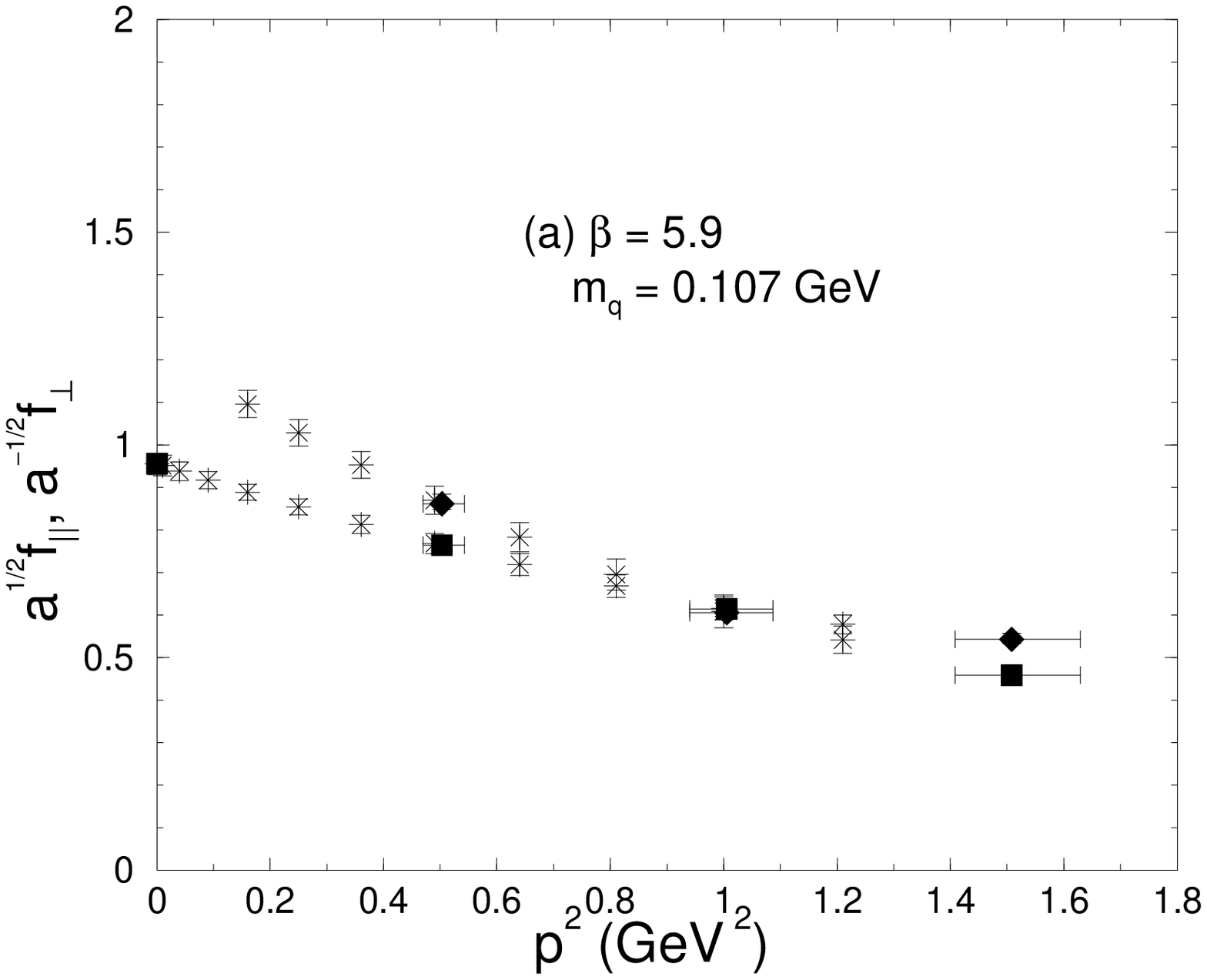}
	\epsfysize=0.45\textheight
		\epsfbox{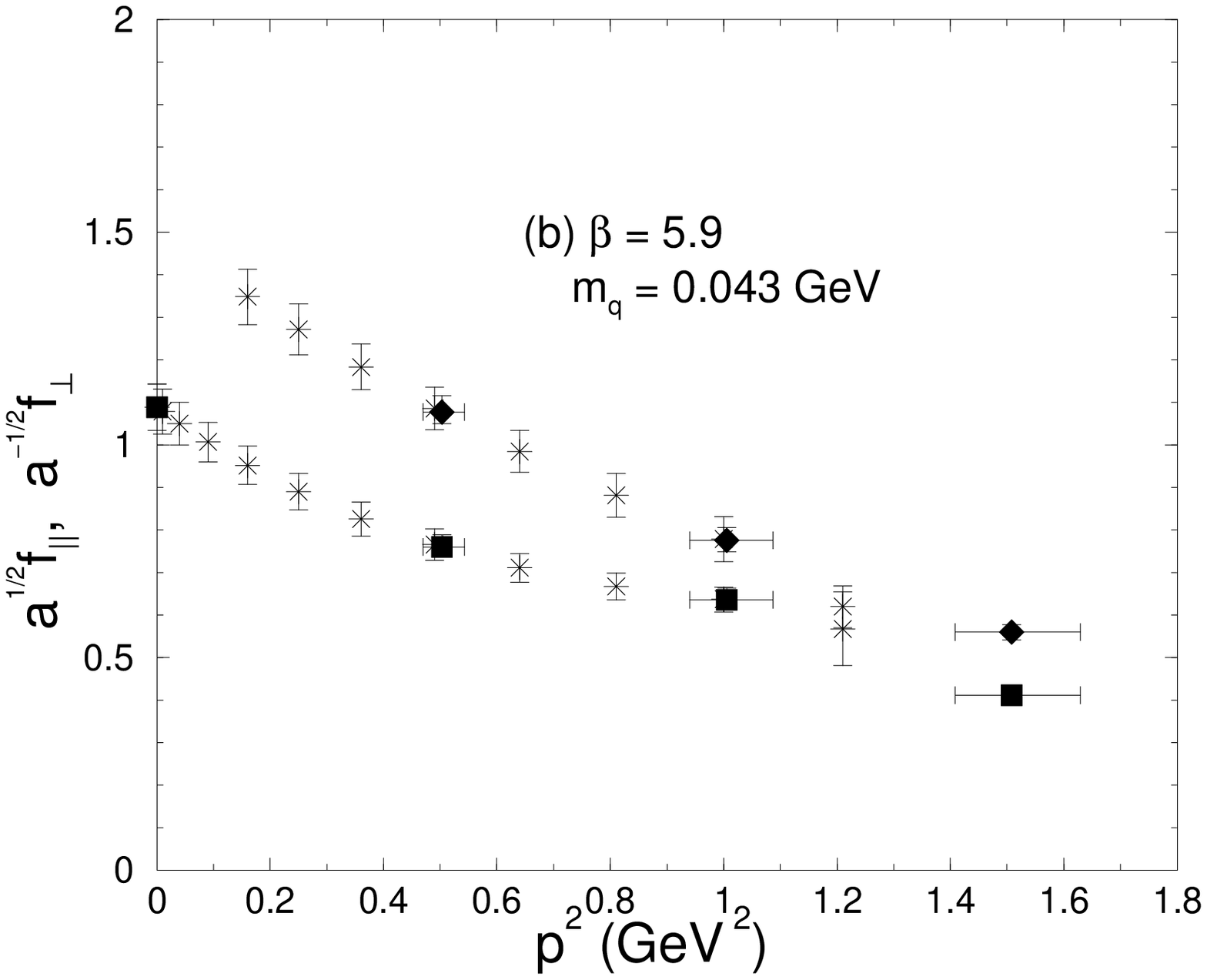}
	\caption[Dpole_effect]{Momentum interpolation (and extrapolation) 
		at $\beta=5.9$, for $D_q\to\eta_q l\nu$ and
		(a) the heaviest of our light quarks, with $\kappa_q = 0.1382$;
		(b) the lightest of the light quarks, with $\kappa_q = 0.1394$.}
	\label{Dpole_effect}
\end{figure}

We must extend the interpolation to an extrapolation to obtain an
estimate of~$f_\perp$ for $p<0.7$~GeV.
As the pion becomes softer and lighter one expects from
Eq.~(\ref{soft_perp}) that the dependence on~$E$ (and hence~$p$)
is sensitive to the Ansatz for extrapolation.
The $B^*$ pole gives $f_\perp$ a peak at low momentum, and the height
of the peak rises as the quark mass decreases.
This shape is hard to capture, as is shown in Fig.~\ref{fig:pole_effect},
unless the fit is constrained to it.
For $p>0.7$~GeV the pole fit agrees perfectly with the method
described above.
But as $p$ is decreased into the region of extrapolation, the two forms
start to deviate.
Above 0.4~GeV the agreement is still good, so we make a cut here.
For smaller momenta phase space suppresses the number of events, so this
cut has no serious ramifications.
For $D$ decays the situation is much the same, as shown in
Fig.~\ref{Dpole_effect}.
Therefore, we impose the same low-momentum cut in this case. 
Other functional forms, such as rational, do not make much difference 
in~$d\Gamma/dp\propto p^4|f_+|^2/E$, once the cut at 
$p_{\text{min}}=0.4$~GeV is imposed.

At high momentum there are other difficulties.
The signal-to-noise ratio of the three-point function deteriorates.
For the highest momentum, $\bbox{n}=(2,0,0)$, we cannot always extract
a convincing matrix element: in some cases the plateau $\Delta t$ is
just 2 time-slices, and three-point functions with different sources
and sinks do not yield the same value for the matrix element.
We cannot include these data in the interpolation.
For the second-highest momentum, $\bbox{n}=(1,1,1)$, we cannot extract
the matrix elements at lighter~$m_q$, so statistical 
errors blow up in the chiral extrapolation.
We therefore place a cut at $\bbox{n}=(1,1,0)$, which corresponds to 
$p_{\text{max}}=1.0$~GeV.
Indeed, our uncertainties would be smaller with a lower upper cut, at 
the cost of reducing the overlap with the experimental data further 
still.

\subsection{Chiral extrapolation}
Following the momentum interpolation, the form factors $f_\parallel$ 
and $f_\perp$ at $\beta=5.7$ and~$5.9$ are extrapolated to the chiral 
limit at fixed momentum, guided by chiral perturbation theory.
From Eqs.~(\ref{soft_parallel}) and (\ref{soft_perp}) one can see that 
the chiral behavior of $f_\parallel$ and $f_\perp$ should be very 
different.
In particular, $f_\parallel$ does not contain a $B^*$ pole, at least
not at the leading order in the chiral expansion.
In the form factors, dependence on the light pseudoscalar mass enters 
both through $m_\pi^2$ and~$E$.
With our momentum cut, $p>0.4$~GeV, and our light meson masses,
$0.45~\text{GeV}<m_{\eta_q}<0.74~\text{GeV}$, the dependence of~$E$
on~$p$ remains smooth, so we try fits of the form
\begin{equation}
	f_{\parallel,\perp} = A + Bm + Cm^2,
	\label{chiral_fits}
\end{equation}
where $m=\log(1+m_0a)$.
We compare quadratic fits with floating~$C$ to linear ones with 
fixed~$C=0$.
The difference in the chiral limit of these different fits is the
origin of our greatest systematic uncertainty.

It would be desirable to have quark propagators at lighter quark 
masses to achieve better control on the chiral extrapolation.
The computer time would increase substantially, however, and the
obstacle of exceptional configurations would have to be overcome, 
for example as in Ref.~\cite{Bardeen:1998gv}.

We note that when $p=0$ (or $p\ll m_{\pi}$) it would be 
better~\cite{Lellouch:1996yv} to carry out the chiral extrapolation at 
fixed~$E$, instead of fixed~$p$.
With $p>0.4$~GeV, however, the fixed~$E$ extrapolation is probably not 
essential, although it may reduce the uncertainty from the chiral 
extrapolation.
We shall investigate this issue elsewhere.

\section{Systematic Errors}
\label{sec:syserr}

As discussed in the previous section, we do not have useful results
outside the range
\begin{equation}
	0.4~\text{GeV}\leq p\leq 1.0~\text{GeV} ,
	\label{range}
\end{equation}
where $p=|\bbox{p}_\pi|$ is the pion's three-momentum in the rest
frame of the $B$ or $D$.
Matrix elements with higher momentum are not estimated reliably,
and at lower momentum the chiral extrapolation used is no longer good.
In this section we analyze the systematic uncertainties 
quantitatively, focusing on the partially integrated
rates~$T_B(0.4~\text{GeV}, 1.0~\text{GeV})$ 
  and~$T_D(0.4~\text{GeV},0.925~\text{GeV})$, 
defined in Eq.~(\ref{T}), and the CKM matrix obtained from 
Eqs.~(\ref{Vub}) and~(\ref{Vcd}).
A~summary of this analysis is given in Table~\ref{tab:budget10}.
\begin{table}
	\centering
	\caption{Budget of statistical and systematic uncertainties
		in this work for the quantities
		$T_B(0.4~\text{GeV},1.0~\text{GeV})$,
		$T_D(0.4~\text{GeV},0.93~\text{GeV})$, and
$T_B(0.4~\text{GeV},0.9~\text{GeV})/T_D(0.4~\text{GeV},0.9~\text{GeV})$.
		All entries in percent.}
	\label{tab:budget10}
	\begin{tabular}{lrrrrrr@{\hspace*{1.5em}}l}
		uncertainty &
		\multicolumn{1}{c}{$~T_B$} & \multicolumn{1}{c}{$|V_{ub}|$} &
		\multicolumn{1}{c}{$~T_D$} & \multicolumn{1}{c}{$|V_{cd}|$} &
		$T_B/T_D$\hspace*{-1.4em} & $|V_{ub}/V_{cd}|$\hspace*{-1.5em} & \\ 
		\hline
		statistical         &\err{27}{9} &\err{14}{5} &\err{17}{8} &\err{9}{4} &\err{10}{4}&\er{5}{2} & \\
		excited states      &   6  &  3  &   6  &  3  &   6  &  3  & \\
		$\bbox{p}$ extrapolation & 10 & 5 &  9  &  5  &   9  &  5  & \\
		$m_q$ extrapolation & \err{16}{22} & \err{8}{11} &
			\err{3}{18} & \er{2}{9} & \err{13}{4} &\er{7}{2} & \\
		adjusting $m_Q$     &   6  &  3  &   2  &  1  &   8  &  4  & \\
		HQET matching       &  10  &  5  &  10  &  5  &  10  &  5  & \\
		$a$ dependence      & \err{16}{3} & \er{8}{2} &
			\err{23}{6} & \err{11}{3} &  5   & 3  & \\
		definition of $a$   &  11  &  6  &   4  &  2  &   8  &  4  & \\
		\hline
		total systematic    & 30 & 15 &
			\err{28}{24} & \err{14}{12} & \err{23}{20} & \err{12}{10} & \\
		total (stat $\oplus$ syst) & \err{40}{31} & \err{20}{16} &
			\err{32}{26} & \err{16}{13} & \err{25}{20} & \err{13}{10} & \\
	\end{tabular}
\end{table}

The statistical error is estimated with the bootstrap method, drawing
1000 samples for each fit.
The bootstrap propagates the statistical uncertainty, including
correlations, through the interpolation in light meson momentum and
extrapolation in light-quark mass, so in the end statistics remain
a quantitatively important source of uncertainty.

\subsection{Excited states}
As explained in Sec.~\ref{sec:numerics}, we take care to isolate the
desired lowest-lying $\pi$ and $B$ states from their radial excitations
when computing the three-point function of Eq.~(\ref{eq:3ptfunc}).
The associated uncertainty on the matrix elements (and, thus, the
form factors) is computed by comparing fits with different smeared
and unsmeared interpolating operators.
After choosing the optimal fit range for each combination of smearing
functions, we find deviations in $F_\perp$ and $F_\parallel$
of 1--3~percent, where the high end of the range is for momenta near 
the upper cut.
We assign an uncertainty of 6~percent to $T_B$ and~$T_D$.

Although we calculate similar matrix elements for each $\bbox{p}$
and for $B$ and $D$ decays, the range~$\Delta t$ of time-slices kept
in the fit was chosen independently for each case.
Therefore, the excited state contamination in $T_B/T_D$ is partly, but
not fully, correlated.
A~conservative error estimate is again 6~percent.

\subsection{Momentum and chiral extrapolations}
\label{subsec:mqp}
The form factor $f_+$ that enters into the partial width
is more sensitive to $f_\perp$ than to~$f_\parallel$.
Thus, it could be sensitive at small~$p$ to the extrapolation described 
in Sec.~\ref{sec:mom}.
The \emph{rate}, however, is much less sensitive,
because phase space suppresses it at small~$p$.
For $T_B$ the variation between linear, rational, and pole forms
is~$\pm 10\%$.

\begin{figure}
\centering
	\epsfysize=0.45\textheight
		\epsfbox{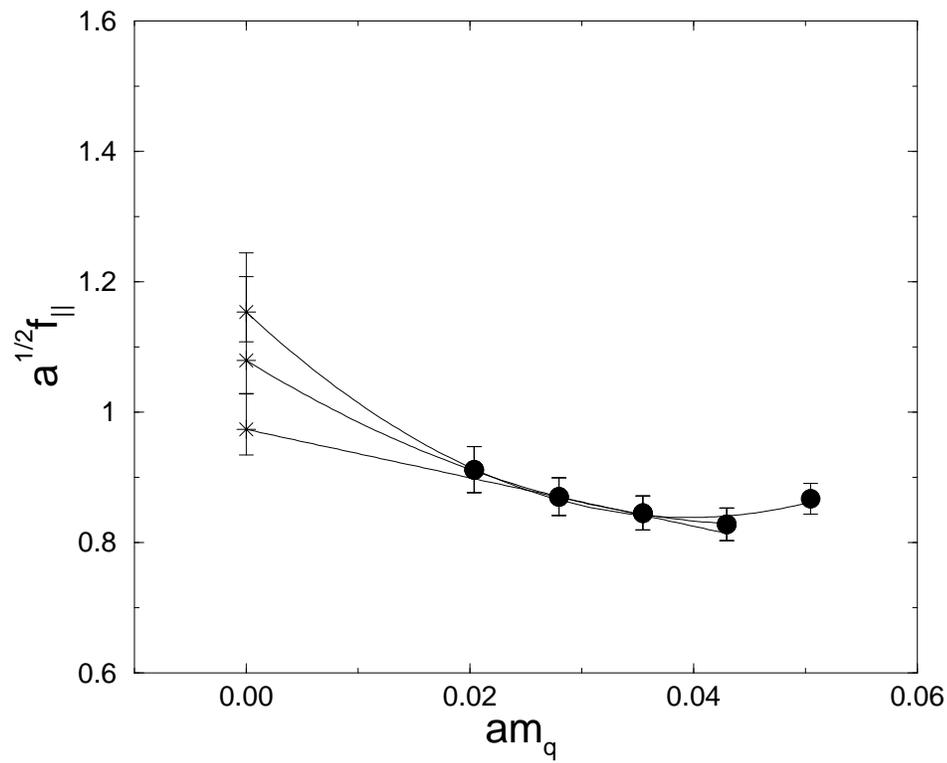}
	\epsfysize=0.45\textheight
		\epsfbox{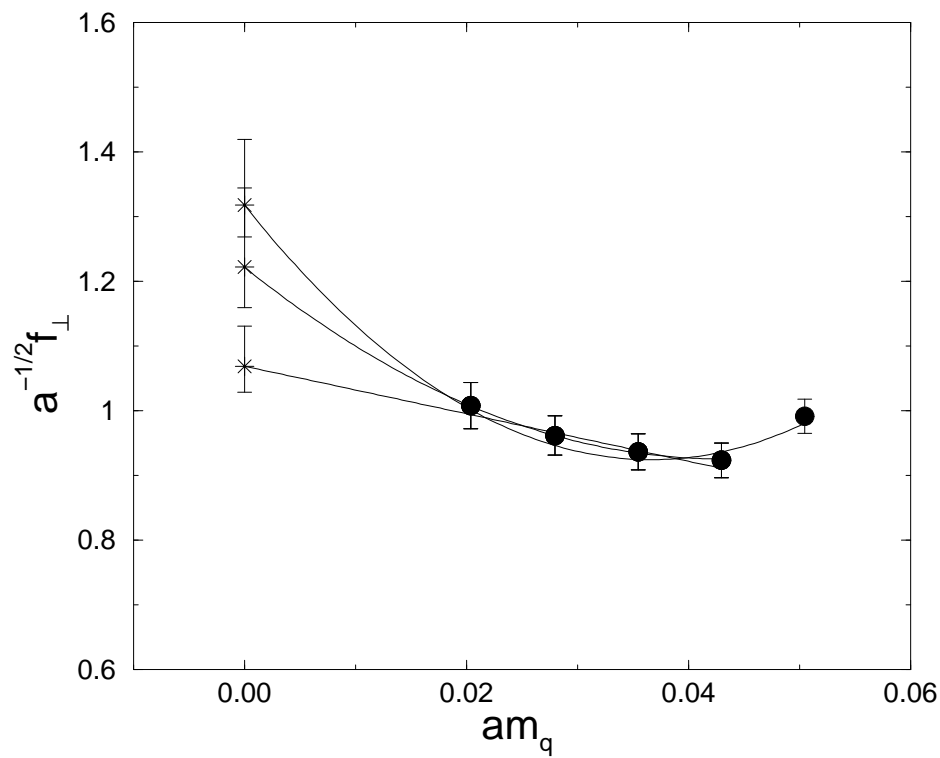}
	\caption[good_extrapsV4]{Chiral extrapolations of
		$f_\parallel$ and $f_\perp$ for $\bbox{n}=(1,0,0)$
		and $\beta=5.9$.}
	\label{good_extrapsV4}
\end{figure}

The chiral extrapolation is a major source of uncertainty.
Figure~\ref{good_extrapsV4} shows the chiral extrapolation at
$\beta=5.9$ for $f_\parallel$ and~$f_\perp$ at $\bbox{n}=(1,0,0)$.
We compare three different fits:
(1) a quadratic fit to the four lightest quark masses;
(2) a linear    fit to the four lightest quark masses; and
(3) a quadratic fit to all five light quark masses.
The first has the lowest $\chi^2/\text{d.o.f.}$, but the other two are
perfectly acceptable.
For other momenta the behavior is the same.
Because the extrapolated result from the first (and best) fit lies
between the other two, we use it to give our central value, and use
the other two as estimates of the systematic error.
The ambiguity of the fits, and hence the systematic error, could be
reduced with explicit calculation at smaller~$m_q$, but a suitable
point is not feasible with our computer resources.
We are left with an uncertainty of
\err{16}{22}\% in~$T_B$ and \err{3}{18}\% in~$T_D$.

The error bars on the extrapolated points in Fig.~\ref{good_extrapsV4}
show how the statistical uncertainties are inflated by the chiral
extrapolation.
This part of the uncertainty is statistical in nature, so it is
incorporated into the first line of Table~\ref{tab:budget10}.
Indeed, it is the main reason the statistical uncertainty in~$T_B$
($T_D$) grows from 6~percent (7~percent) with $m_q=m_s$ to 18~percent
(13~percent) with $m_q=m_d$.

\subsection{Heavy quark mass dependence}
\begin{figure}[bp]
	\centering
    \epsfysize=0.45\textheight
		\epsfbox{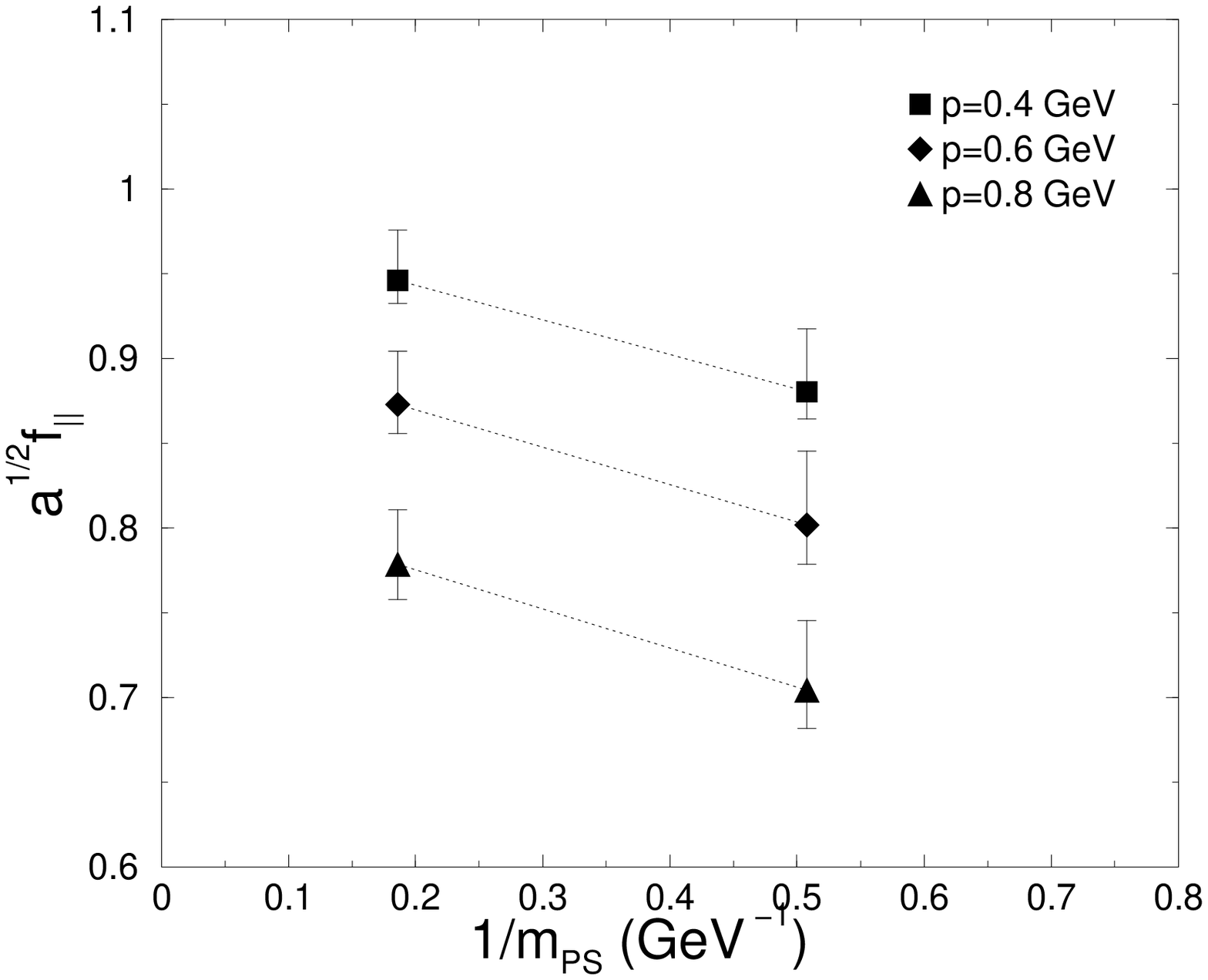}
    \epsfysize=0.45\textheight
		\epsfbox{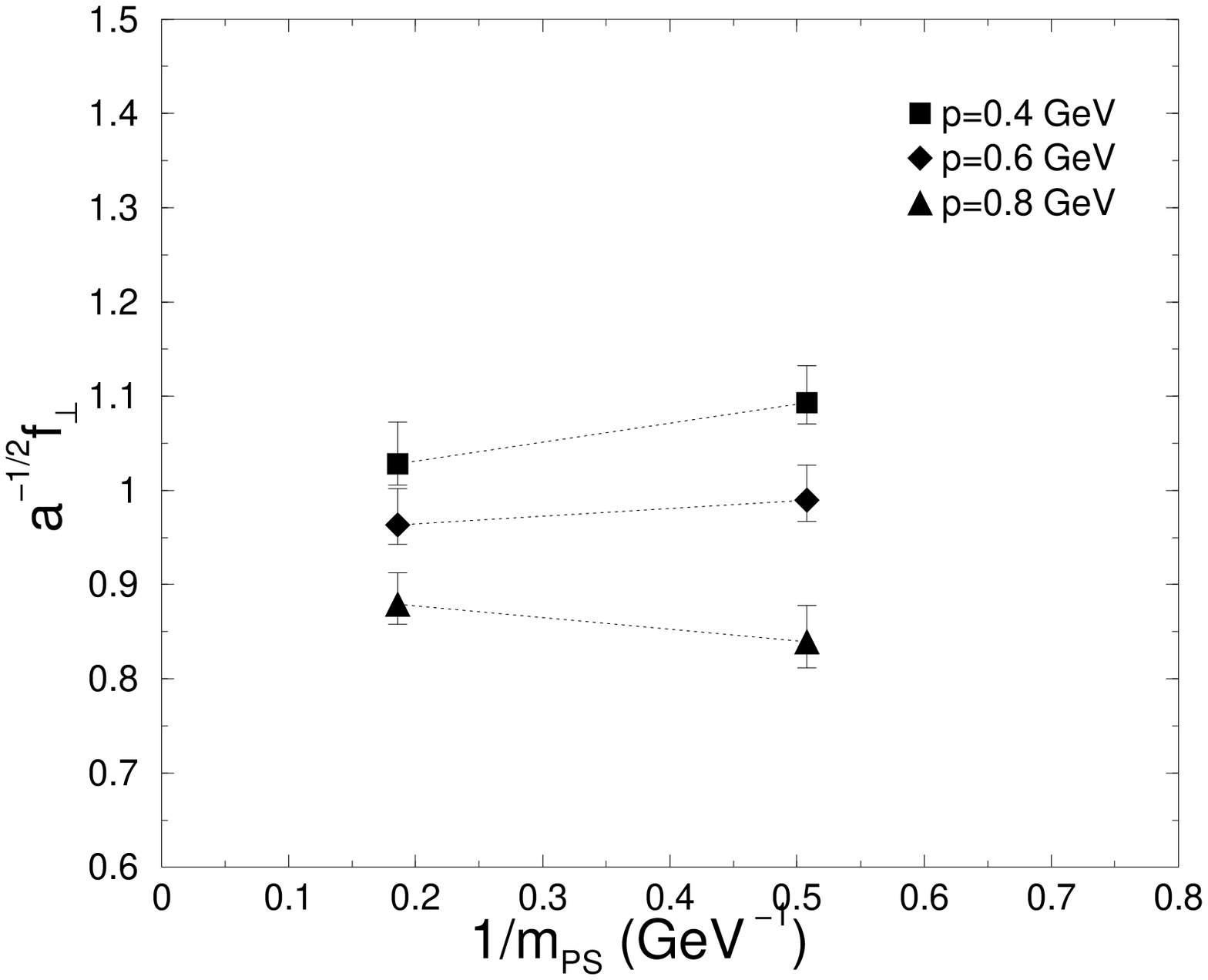}
    \caption[hq_dependence]{The heavy quark mass dependence at
		several momenta, at $\beta=5.9$ for the temporal matrix
		element.  The light quarks are strange quarks.
		The dotted lines are to guide the eye.}
	\label{hq_dependence}
\end{figure}

To examine the dependence on the heavy quark mass we use form factors 
with a light strange quark, because then statistical errors do not 
mask the effect.
Figure~\ref{hq_dependence} compares the form factors $B_s$ and $D_s$
decays.
There is a significant difference.
The quarkonium spectrum tunes the (bare) heavy quark mass within a
precision of 1--2\%~\cite{El-Khadra:1998hq}, which clearly would have
no significant effect on the form factors.
But because of lattice artifacts in the quarkonium binding
energy~\cite{Kronfeld:1997uy} and because of quenching, the heavy-light
spectrum yields a different adjustment of bare quark masses.
The shift is to lower $1/m_{\text{PS}}$ in Fig.~\ref{hq_dependence}.
From Eq.~(\ref{f+}) one sees that $f_\perp$ dominates in~$f_+$ for
$B$ decay.
Thus, $f_+$ is smaller with the heavy-light adjustment of the bottom
quark mass, and $T_B$ is 6~percent smaller.
On the other hand, $f_\perp$ and~$f_\parallel$ make a comparable
contributions to $f_+$ for $D$ decay.
It turns out that~$f_+$ is larger with the heavy-light adjustment of
the charmed quark mass, and $T_D$ is 2~percent larger.
The ratio $T_B/T_D$ is 8~percent smaller.

\subsection{Matching}
\label{matching}
As explained in Sec.~\ref{sec:bckgnd}, our treatment of the heavy quark
matches lattice gauge theory with Wilson fermions to HQET.
This requires calculations of the short-distance coefficients:
$1/m_2$ and $1/m_{\cal B}$ in the effective action;
$\sqrt{Z_{V^{uu}}Z_{V^{bb}}}$ in the
definition of the current; and $\rho_{V_\parallel}$, $\rho_{V_\perp}$,
and $1/m_3$ in the description of the currents.
As discussed in the previous subsection, $m_2$ is adjusted
non-perturbatively, by tuning the quarkonium spectrum to agree with
experiment.
The normalization factors $Z_{V^{uu}}$ and $Z_{V^{bb}}$ are also 
computed non-perturbatively, by requiring that flavor-conserving 
matrix elements $\pi\to\pi$ and $B\to B$, computed by analogy with 
Eq.~(\ref{three-pt}), give unit charge.
The uncertainty from it is purely statistical and much smaller than
all other statistical uncertainties.

The significant systematic effects in the matching procedure come
from computing $\rho_{V_\parallel}$ and $\rho_{V_\perp}$, and from the
mismatch between Eqs.~(\ref{VlatHQET}) and Eqs.~(\ref{VcontHQET})
at the level of dimension-four and higher currents.
In the present work we do this part of the matching with perturbative 
QCD, leading to errors of order $\alpha_s^2$, $\alpha_s/m_Q$,
$1/m_Q^2$, respectively.
Let us now consider these effects in turn.

Because they are short-distance quantities, the matching factors
$\rho_{V_\parallel}$ and $\rho_{V_\perp}$ should be calculable in
perturbation theory.
(Note that all effects that make lattice perturbation theory less
reliable than continuum perturbation theory are absorbed
into~$\sqrt{Z_{V^{uu}}Z_{V^{bb}}}$.)
We have calculated them to one loop, so we write
\begin{equation}
	\rho_V = 1 + \alpha_s(q^*) 4\pi \rho_V^{[1]}
\end{equation}
for $\rho_{V_\parallel}$ and $\rho_{V_\perp}$.
We use the Brodsky-Lepage-Mackenzie (BLM) procedure to choose the
expansion parameter
$\alpha_s(q^*)$~\cite{Brodsky:1983gc,Lepage:1993xa}.
In the scheme in which the Fourier transform of the heavy-quark
potential reads $V(k)=-C_F4\pi\alpha_s(k)/k^2$, the BLM scale $q^*$ is
given through
\begin{equation}
	\log(q^*a) = \frac{{}^*\rho_V^{[1]}}{2\rho_V^{[1]}}
\end{equation}
where ${}^*\rho_V^{[1]}$ is obtained from $\rho_V^{[1]}$ by replacing
the gluon propagator~$D(k)$ with $D(k)\log{k^2a^2}$.
The details of these calculations are similar to those described
in Ref.~\cite{Kronfeld:1999tk}, and the results are listed in
Table~\ref{tab:PT}~\cite{Harada}.
\begin{table}
	\centering
	\caption{Perturbation theory for matching factors
		$\rho_{V_\parallel}$ and $\rho_{V_\perp}$.
		The one-loop terms $\rho_V^{[1]}$ and $^*\rho_V^{[1]}$
		are in units of $10^{-5}$.}
	\label{tab:PT} 
	\begin{tabular}{crrccrrccrrcc}
		&  \multicolumn{4}{c}{$\beta=6.1$} &
	\multicolumn{4}{c}{$\beta=5.9$} & \multicolumn{4}{c}{$\beta=5.7$} \\
	    & 
	$\rho_V^{[1]}$ & $^*\rho_V^{[1]}$ & $\alpha_s(q^*)$ & $\rho_V$ &
	$\rho_V^{[1]}$ & $^*\rho_V^{[1]}$ & $\alpha_s(q^*)$ & $\rho_V$ &
	$\rho_V^{[1]}$ & $^*\rho_V^{[1]}$ & $\alpha_s(q^*)$ & $\rho_V$ \\
	\hline
	$b$ & & & & & & & & & & & & \\
	$f_\parallel$ &
	   536  &    980  & 0.159 & 1.011 &
	   817  &   1591  & 0.173 & 1.018 &
	  1065  &   2199  & 0.196 & 1.026 \\
	$f_\perp$     & 
	$-1987$ & $-3312$ & 0.163 & 0.959 &
	$-2096$ & $-3534$ & 0.181 & 0.952 &
	$-2146$ & $-3621$ & 0.212 & 0.943 \\
	\hline
	$c$ & & & & & & & & & & & & \\
	$f_\parallel$ & 
	$-  59$ &     13  & 0.233 & 0.998 &
	$-  28$ &     40  & 0.402 & 0.999 &
	$+  63$ &    152  & 0.184 & 1.001 \\
	$f_\perp$     & 
	$- 947$ & $-1368$ & 0.169 & 0.980 &
	$-1223$ & $-1821$ & 0.188 & 0.971 &
	$-1508$ & $-2339$ & 0.218 & 0.959 \\
	\end{tabular}
\end{table} 

The effects are small for $B$ decays and tiny for $D$ decays.
This can be understood because the $\rho_V$s are ratios of very
similar quantities, so there is good cancellation.
It is therefore plausible that the two-loop contribution is numerically
smaller by another factor of $\alpha_s\approx 0.2$, and thus completely
negligible.

Next, we must estimate the uncertainty from the mismatch of the
$1/m_Q$ term in the heavy-quark expansion.
This contributes an error on either form factor~$f$
\begin{equation}
	\delta_{1/m_Q} f \sim \alpha_s b_{1/m_Q}(m_Qa) 
		m_Q^{-1}\Lambda_{\text{QCD}} f
	\label{eq:mQ1}
\end{equation}
from $1/m_3$ and $1/m_{\cal B}$ contributions, and $b_{1/m_Q}$ gives
the deviation of the short-distance coefficients for the lattice and
continuum theories.
(See Refs.~\cite{El-Khadra:1997mp,Kronfeld:2000ck} for further details.)
The factor~$b(m_Qa)$ is at most of order unity; for our calculations of
$D$-meson matrix elements it is of order $m_ca<1$.
Taking $\alpha_s\approx 0.2$ and $\Lambda_{\text{QCD}}\approx 500$~MeV
one finds that these errors, in either case, are at most a few percent
on~$f$ or the CKM matrix.

Finally, we must estimate the uncertainty from the mismatch of the
$1/m_Q^2$ terms:
\begin{equation}
	\delta_{1/m_Q^2} f \sim b_{1/m_Q^2}(m_Qa) 
		m_Q^{-2}\Lambda_{\text{QCD}}^2 f.
	\label{eq:mQ2}
\end{equation}
There are many contribution at order~$1/m_Q^2$ in the heavy-quark
expansion, most of which come from iteration of the $1/m_Q$ terms.
Only genuine $1/m_Q^2$ terms in the effective action and currents
can be as inaccurate as Eq.~(\ref{eq:mQ2}) suggests.
Since $\Lambda_{\text{QCD}}/m_Q\approx\Lambda_{\text{QCD}}a%
\lesssim\alpha_s$ for our lattice data, the error $\delta_{1/m_Q^2}f$
is similar in magnitude to that of~$\delta_{1/m_Q}f$.

The estimates in Table~\ref{tab:budget10} derived from 
Eqs.~(\ref{eq:mQ1}) and~(\ref{eq:mQ2}) are very conservative.
It is plausible that the denominator of heavy-quark expansion is 
$2m_Q$, and it is possible that the unknown coefficients are fractions 
instead of 1--2 as used above.
Thus, the matching uncertainties may already be negligible.

The masses of the $b$ and $c$ quarks differ by about a factor of three.
The short-distance coefficients are functions of
$m_Qa$~\cite{El-Khadra:1997mp,Kronfeld:2000ck}, so the matching
uncertainties do not cancel completely in the ratio~$T_B/T_D$.
In particular, on our lattices the mismatch coefficients $b_{1/m_Q^n}$
are of order~1 for $b$ quarks, but $b_{1/m_Q}\sim m_ca$ and
$b_{1/m_Q^2}\sim (m_ca)^2,\,\alpha_s m_ca$ for $c$~quarks.
Nevertheless, the effects often have the same sign, so we take the
uncertainty in the ratio to be the same as in numerator or denominator.

\subsection{Lattice spacing dependence}
For the artificial decays $B_s, D_s\to\eta_s l\nu$ we have results
at three lattice spacings, so we can examine how severely the form
factors are affected.
These decays are good for studying the $a$ dependence, because their 
form factors have small statistical errors.
After chiral extrapolation, on the other hand, the larger statistical 
error bars would mask lattice spacing effects. 
Previous experience with decay constants~\cite{El-Khadra:1998hq}
leads us to believe this will not change very much after chiral
extrapolation.
With the action used in this work the lattice spacing dependence is a
combination of $O(\alpha_sa)$ and $O(a^2)$ effects from
the light quarks and gluons, and the $a$ dependence of the heavy-quark
short-distance coefficients, discussed in the previous subsection.
In particular, when the $\eta_s$ has non-zero recoil momentum~$p$, the 
light-quark lattice effects are $O(\alpha_spa)$ and~$O(p^2a^2)$.

\begin{figure}
	\centering
	\epsfysize=0.45\textheight
		\epsfbox{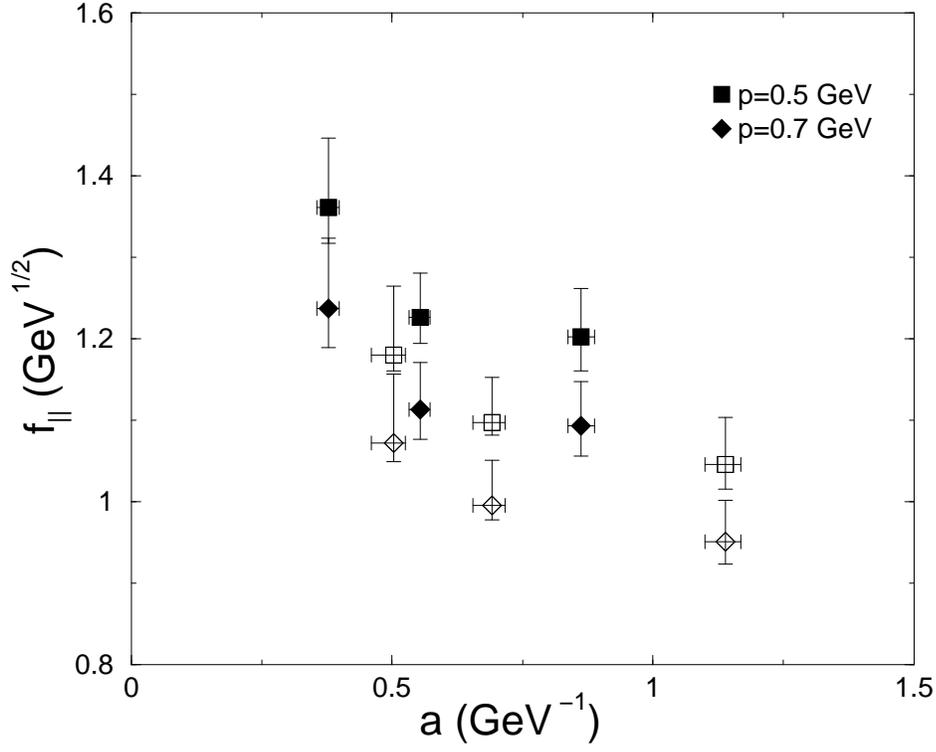}
	\epsfysize=0.45\textheight
		\epsfbox{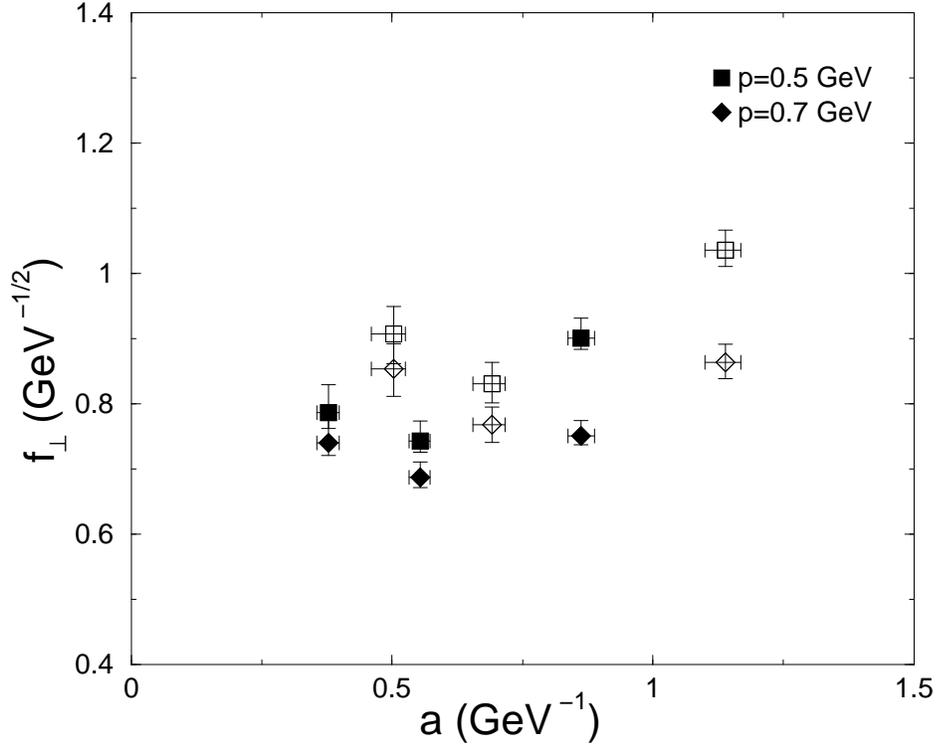}
		\caption[fig:conextraps]{Lattice spacing dependence of 
		$B_s\to\eta_sl\nu$ form factors 
		(a) $f_\parallel$ and (b) $f_\perp$,
		for $p=0.5$~GeV (squares) and $0.7$~GeV (diamonds).
		The solid (open) symbols correspond to defining the lattice
		spacing with $\Delta m_{\text{1P-1S}}$ ($f_\pi$).}
	\label{fig:conextraps}
\end{figure}

The $a$ dependence of the form factors is shown 
in Fig.~\ref{fig:conextraps}.
The variation with $a$ is several percent, which is comparable to the
statistical uncertainty and also to the errors from the mismatch of
the heavy quark.
The observed $a$ dependence is therefore a combination of
(uncorrelated) statistical fluctuations, lattice artifacts from the
light degrees of freedom, and from the lattice artifacts [described
in Eqs.~(\ref{eq:mQ1}) and~(\ref{eq:mQ2})] of the heavy quark.
They cannot be disentangled with the current set of calculations,
so it does not make sense to extrapolate~$a\to 0$.

\begin{figure}
	\centering
	\epsfysize=0.45\textheight 
		\epsfbox{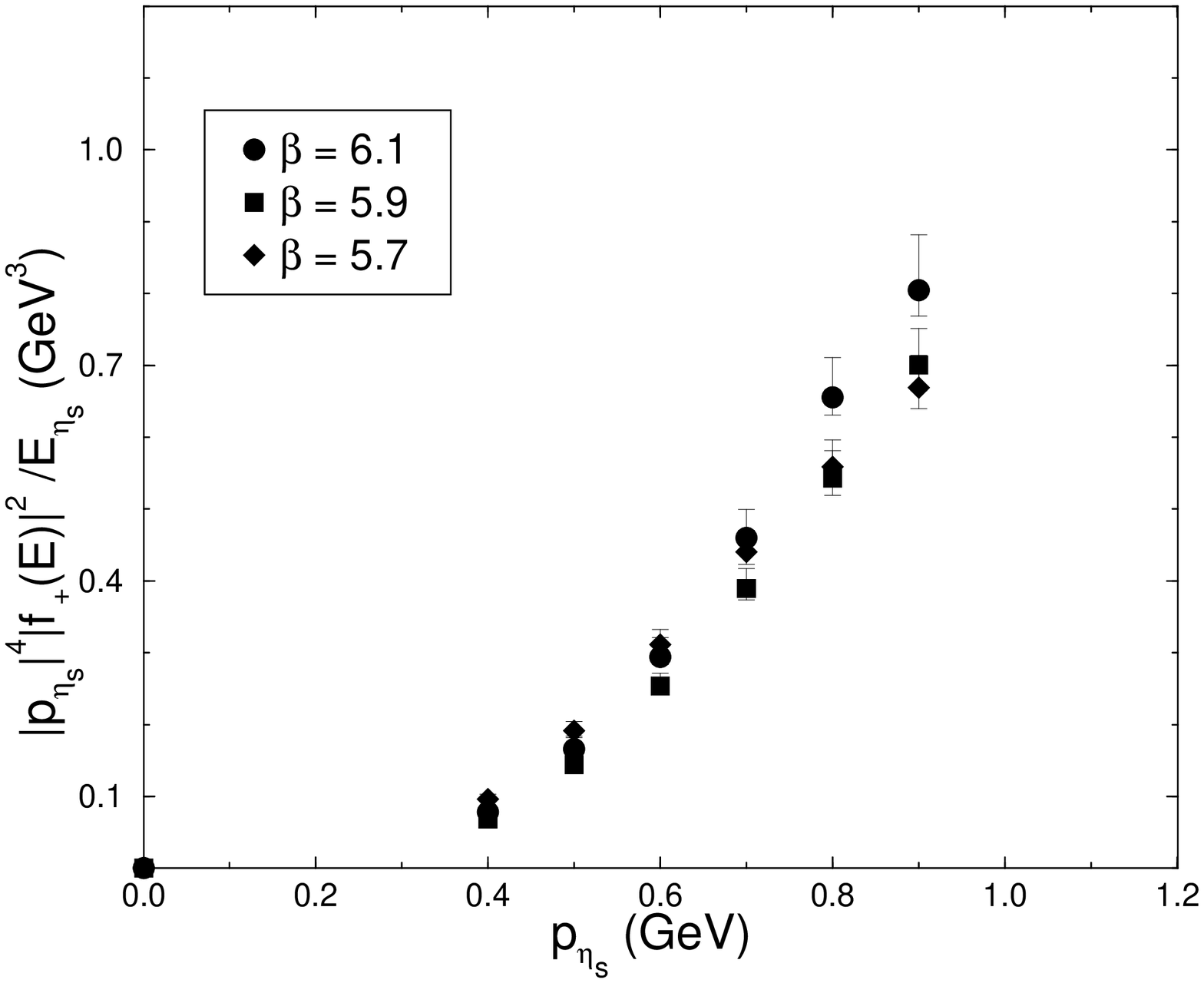}
	\epsfysize=0.45\textheight 
		\epsfbox{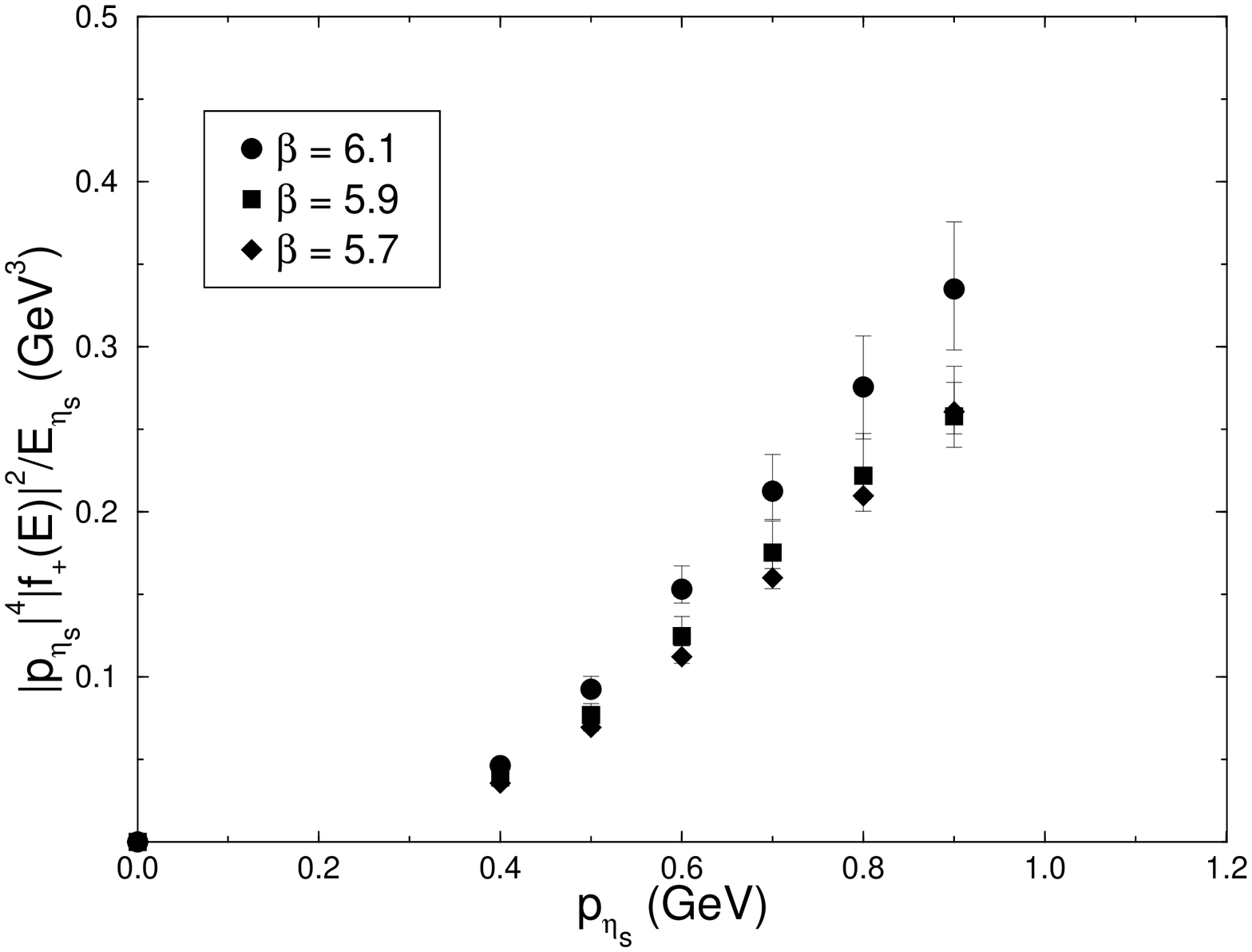}
    \caption[fig:scalingDGamma_dpD]{Comparison of $p^4|f_+|^2/E$ at
		three lattice spacings for
		(a) $B_s\to\eta_sl\nu$ and (b) $D_s\to\eta_sl\nu$.}
	    \label{fig:scalingDGamma_dpD}
\end{figure}

Instead we choose the results from $\beta=5.9$, where we have the
widest range of light quark masses, for our central value and use
the other two lattices to estimate the uncertainty.
Figure~\ref{fig:scalingDGamma_dpD} shows the combination
$p^4|f_+|^2/E$, which is proportional to $d\Gamma/dp$, at all three
lattice spacings for $B_s\to \eta_sl\nu$ and $D_s\to \eta_sl\nu$.
As in Fig.~\ref{fig:conextraps} one sees that the dependence on~$a$ is
several percent, and increases with increasing~$p$.
By integrating over $p$ we find a variation of of~$\err{16}{3}\%$
in $T_B$ and~$\err{23}{6}\%$ in~$T_D$.

\subsection{Definition of $a$ and quenching}
Changes in the final results from changing the definition of~$a$
can be thought of as a crude way to estimate effects of the quenched
approximation.
In lattice units we obtain $f_\parallel a^{1/2}$ and
$f_\perp a^{-1/2}$, so converting to physical units introduces a mild
explicit dependence on the value chosen for~$a$.
There is also an implicit dependence that enters through functional
dependence on~$E$ (or~$p$).
These two effects are illustrated in Fig.~\ref{fig:conextraps}.
The solid (open) points are obtained by defining~$a$ so that the 1P-1S
splitting of charmonium (pion decay constant) takes its physical value.
The central values in the paper are computed with the 1P-1S definition.
At $\beta=5.9$ we repeat the full analysis with the $f_\pi$ definition.
We find that $T_B$ ($T_D$) increases by 11~percent (4~percent), and
the ratio $T_B/T_D$ increases by 8~percent.

A more serious estimate of the effect of the quenched approximation is
impossible without generating gauge fields with dynamical quark loops.
This would require more computer resources than we have at our disposal,
and no other group has yet studied these semileptonic decays with
dynamical quarks.
There are results with two light, dynamical flavors for the leptonic 
decay constants~$f_B$ and~$f_D$, using either 
lattice NRQCD~\cite{AliKhan:2000zr} 
or our method~\cite{AliKhan:2000eg,Bernard:2000nv} for the heavy quark.
In that case one finds an increase of between 10--11 percent over the 
quenched result.

The exercise of changing the definition of~$a$ easily could underestimate
the effects of quenching.
At the same time, we do not expect form factors to be more sensitive
than~$f_B$.
Thus, a provisional estimate of a uncertainty in $T_{B,D}$ of 10--20\%
seems reasonable.

\section{Results}
\label{sec:results}

The main results of this paper, given in Eqs.~(\ref{TB10}) 
and~(\ref{TD09}), are the quantities
$T_B(0.4~\text{GeV},1.0~\text{GeV})$ and
$T_D(0.4~\text{GeV},0.925~\text{GeV})$, which are proportional to the 
partially integrated rates.
It may also be of interest to present the results in other ways.
In this section we give results for the ratio $T_B/T_D$, as well as 
results for $T_B$, $T_D$, and $T_B/T_D$ with a lower upper cut.
We also give results for the form factors themselves.

Many uncertainties cancel in the ratio of $B$ and $D$~rates:
the statistical error is correlated, and the systematic errors are
similar in nature.
Because of heavy-quark symmetry it is most sensible to form a ratio
with the same cuts for both.
We find
\begin{equation}
\frac{T_B(0.4~\text{GeV},0.9~\text{GeV})}%
{T_D(0.4~\text{GeV},0.9~\text{GeV})}   =   2.04^{+\,0.20}_{-\,0.09}
		{\;}^{+\,0.26}_{-\,0.08}		
		\pm 0.10						
		\pm 0.20						
		\pm 0.29,						
\end{equation}
where the uncertainties are from statistics, chiral extrapolation, 
$a$~dependence, HQET matching, and other miscellaneous sources.
A~more detailed budget of the last uncertainty is given in 
Table~\ref{tab:budget10}.

As mentioned above, raising the upper cut $p_{\text{max}}$ increases 
the uncertainty.
Conversely, lowering $p_{\text{max}}$ decreases the uncertainty.
Repeating the full analysis at $p_{\text{max}}=0.8~\text{GeV}$,
we find
\begin{eqnarray}
	T_B(0.4~\text{GeV},0.8~\text{GeV}) & = & 0.294^{+\,0.063}_{-\,0.031}
		{\;}^{+\,0.041}_{-\,0.064}		
		{\,}^{+\,0.041}_{-\,0.006}		
		\pm 0.029						
		\pm 0.038~\text{GeV}^4,			
	\label{TB08} \\
	T_D(0.4~\text{GeV},0.8~\text{GeV}) & = & 0.145^{+\,0.026}_{-\,0.013}
		{\;}\pm 0.016					
		{\,}^{+\,0.024}_{-\,0.012}		
		\pm 0.014						
		\pm 0.017~\text{GeV}^4.			
	\label{TD08}
\end{eqnarray}
and the ratio
\begin{equation}
	\frac{T_B(0.4~\text{GeV},0.8~\text{GeV})}%
{T_D(0.4~\text{GeV},0.8~\text{GeV})}    =    2.03^{+\,0.19}_{-\,0.10}
		{\;}^{+\,0.25}_{-\,0.16}		
		\pm 0.10						
		\pm 0.20						
		\pm 0.24.						
\end{equation}
Table~\ref{tab:budget08} shows a budget of systematic errors, similar to
Table~\ref{tab:budget10}.
\begin{table}
	\centering
	\caption{Budget of statistical and systematic uncertainties
		in this work for the quantities
		$T_B(0.4~\text{GeV},0.8~\text{GeV})$,
		$T_D(0.4~\text{GeV},0.8~\text{GeV})$, and
      $T_B(0.4~\text{GeV},0.8~\text{GeV})/T_D(0.4~\text{GeV},0.8~\text{GeV})$.
		All entries in percent.}
	\label{tab:budget08}
	\begin{tabular}{lrrrrrr@{\hspace*{1.5em}}l}
		uncertainty &
		\multicolumn{1}{c}{$~T_B$} & \multicolumn{1}{c}{$|V_{ub}|$} &
		\multicolumn{1}{c}{$~T_D$} & \multicolumn{1}{c}{$|V_{cd}|$} &
		$T_B/T_D$\hspace*{-1.4em} & $|V_{ub}/V_{cd}|$\hspace*{-1.5em} & \\ 
		\hline
		statistical         &\err{21}{10}&\err{11}{5} &\err{18}{9}&\err{9}{5} &\err{10}{5} &\err{5}{3} & \\
		excited states      &   4  &  2  &   4  &  2  &   4  &  2  & \\
         	$\bbox{p}$ extrapolation & 8 & 4 &   10  &  5  &  6   & 3 & \\
		$m_q$ extrapolation & \err{14}{22} & \err{7}{11} &
			\err{11}{11} & \er{6}{6} &\err{13}{8} &\er{7}{4} & \\
		adjusting $m_Q$     &   3  &  1  &   4  &  2  &   6  &  3  & \\
		HQET matching       &  10  &  5  &  10  &  5  &  10  &  5  & \\
		$a$ dependence      & \err{14}{2} & \er{7}{1} &
			\err{17}{8} & \er{9}{4} &  5   &  3  & \\
		definition of $a$   &   9  &  5  &   3  &  2  &  8   & 4   & \\
		\hline
		total systematic    & \err{26}{28} & \err{13}{14} &
			\err{26}{21} & \err{13}{10} & \err{21}{19} & \err{11}{9} & \\
		total (stat $\oplus$ syst) & \err{33}{29} & \err{17}{15} &
			\err{31}{23} & \err{16}{11} & \err{23}{20} & \err{12}{9} & \\
	\end{tabular}
\end{table}

As one can see from comparing the last two lines in Tables~\ref{tab:budget10} 
and~\ref{tab:budget08}, the total uncertainty is several percent lower
with~$p_{\text{max}}=0.8~\text{GeV}$.

Finally, we give our results for the form factors.
Table~\ref{tab:B} gives the results for form factors in the decay
$B\to\pi l\nu$.
\begin{table}
\centering
\caption[tab:B]{Form factors with statistical and total systematic
errors for the decay $B\to\pi l\nu$.}
\label{tab:B}
\begin{tabular}{ccc|ccccc}
$p$ &  $E$  &  $q^2$  &       $f_{\parallel}$       &         $f_{\perp}$          &             $f_+$            &            $f_0$            &  $p^4|f_+|^2/E$  \\
GeV &  GeV  & GeV$^2$ &         GeV$^{1/2}$         &         GeV$^{-1/2}$         &                              &                             &     GeV$^3$      \\
\hline
0.0 & 0.140 &  26.41  & 1.93\err{29}{3}\err{28}{28} &                              &                              & 1.17\err{18}{6}\err{18}{18} & 0.0 \\
0.1 & 0.172 &  26.07  & 1.88\err{28}{3}\err{28}{28} &                              &                              &                             &                  \\
0.2 & 0.244 &  25.31  & 1.85\err{27}{3}\err{28}{28} &                              &                              &                             &                  \\
0.3 & 0.331 &  24.39  & 1.80\err{25}{4}\err{27}{27} &                              &                              &                             &                  \\
0.4 & 0.424 &  23.41  & 1.73\err{23}{4}\err{26}{26} & 1.05\err{16}{18}\err{16}{16} & 2.10\err{29}{25}\err{32}{32} & 1.00\err{13}{3}\err{15}{15} & 0.27\er{8}{6}    \\
0.5 & 0.519 &  22.41  & 1.65\err{21}{6}\err{25}{25} & 0.99\err{13}{14}\err{15}{15} & 1.96\err{24}{20}\err{29}{29} & 0.95\err{12}{3}\err{14}{14} & 0.46\err{12}{9}  \\ 
0.6 & 0.616 &  21.38  & 1.56\err{20}{7}\err{23}{23} & 0.95\err{10}{9}\err{14}{14}  & 1.84\err{20}{14}\err{27}{27} & 0.89\err{10}{4}\err{13}{13} & 0.71\err{16}{10} \\
0.7 & 0.714 &  20.35  & 1.45\err{18}{7}\err{22}{22} & 0.91\er{9}{5}\err{14}{14}    & 1.72\err{18}{8}\err{26}{26}  & 0.83\err{10}{4}\err{12}{12} & 1.00\err{22}{9}  \\
0.8 & 0.812 &  19.31  & 1.34\err{17}{8}\err{20}{20} & 0.86\err{12}{4}\err{13}{13}  & 1.59\err{21}{7}\err{24}{24}  & 0.76\err{10}{4}\err{11}{11} & 1.27\err{36}{11} \\
0.9 & 0.911 &  18.27  & 1.23\err{17}{7}\err{18}{18} & 0.73\err{15}{6}\err{11}{11}  & 1.36\err{23}{9}\err{20}{20}  & 0.70\er{9}{4}\err{11}{11}   & 1.34\err{51}{17} \\
1.0 & 1.01  &  17.23  & 1.15\err{16}{6}\err{17}{17} & 0.59\err{15}{6}\er{9}{9}     & 1.13\err{24}{9}\err{17}{17}  & 0.64\er{9}{3}\err{10}{10}   & 1.30\err{60}{35} \\
\end{tabular}
\end{table}

%
Listed are $f_\parallel$ and $f_\perp$, which emerge directly from
our lattice calculations, and $f_+$ and $f_0$, which appear in the
expression for the differential rate.
In every case the first error is statistical and the second adds
the systematic uncertainties in quadrature.
Table~\ref{tab:D} lists the same information for $D\to\pi l\nu$.
\begin{table}
\centering
\caption[tab:D]{Form factors with statistical and total systematic
errors for the decay $D\to\pi l\nu$.}
\label{tab:D}
\begin{tabular}{ccc|ccccc}
$p$ &  $E$  &  $q^2$  &       $f_{\parallel}$       &         $f_{\perp}$          &             $f_+$            &            $f_0$            & $p^4|f_+|^2/E$ \\
GeV &  GeV  & GeV$^2$ &         GeV$^{1/2}$         &         GeV$^{-1/2}$         &                              &                             &    GeV$^3$     \\
\hline
0.0 & 0.140 &   2.99  & 1.34\err{19}{3}\err{17}{15} &                              &                              & 1.29\err{20}{2}\err{17}{14} & 0.0              \\ 
0.1 & 0.172 &   2.87  & 1.33\err{19}{2}\err{17}{15} &                              &                              &                             &                  \\ 
0.2 & 0.244 &   2.60  & 1.32\err{18}{2}\err{17}{14} &                              &                              &                             &                  \\ 
0.3 & 0.331 &   2.28  & 1.31\err{17}{3}\err{17}{14} &                              &                              &                             &                  \\ 
0.4 & 0.424 &   1.93  & 1.28\err{16}{4}\err{16}{14} & 1.19\err{16}{15}\err{15}{13} & 1.56\err{17}{10}\err{20}{17} & 1.14\err{13}{4}\err{15}{12} & 0.15\er{3}{2}    \\ 
0.5 & 0.519 &   1.57  & 1.21\err{15}{5}\err{16}{13} & 1.17\err{12}{12}\err{15}{13} & 1.45\err{14}{8}\err{19}{16}  & 1.08\err{12}{4}\err{14}{12} & 0.25\er{5}{3}    \\ 
0.6 & 0.616 &   1.21  & 1.12\err{14}{6}\err{15}{12} & 1.13\err{10}{8}\err{15}{12}  & 1.32\err{12}{6}\err{17}{14}  & 1.01\err{11}{5}\err{13}{11} & 0.36\er{7}{3}    \\ 
0.7 & 0.714 &   0.85  & 1.04\err{13}{6}\err{14}{11} & 1.08\er{7}{6}\err{14}{12}    & 1.18\err{10}{6}\err{15}{13}  & 0.96\er{9}{5}\err{12}{10}   & 0.47\er{8}{4}    \\ 
0.8 & 0.812 &   0.478 & 0.99\err{11}{6}\err{13}{11} & 1.02\er{8}{6}\err{13}{11}    & 1.07\er{9}{6}\err{14}{11}    & 0.95\er{9}{5}\err{12}{10}   & 0.58\err{11}{6}  \\ 
0.9 & 0.911 &   0.109 & 0.95\err{10}{5}\err{12}{10} & 0.98\er{8}{7}\err{13}{11}    & 0.98\er{9}{6}\err{13}{11}    & 0.94\er{9}{5}\err{12}{10}   & 0.69\err{13}{8}  \\
0.925 & 0.935 &   0   & 0.93\err{13}{4}\err{12}{10} & 0.95\er{9}{7}\err{12}{10}    & 0.94\err{10}{5}\err{12}{10}  & 0.94\err{10}{6}\err{12}{10} & 0.71\err{19}{9}  \\
\end{tabular}
\end{table}

%
Our final results for $f_\parallel$ and $f_\perp$ are obtained at
$\beta=5.9$, after chiral extrapolation, with the systematic errors
estimated as described in Sec.~\ref{sec:syserr}.
In particular, the estimate of lattice spacing effects uses results
from all three lattice spacings.
For $p<0.4$~GeV our extrapolation of $f_\perp$ in the pion momentum
is no longer reliable, so we do not quote results for it.

\begin{figure}
	\centering
	\epsfysize=0.45\textheight 
		\epsfbox{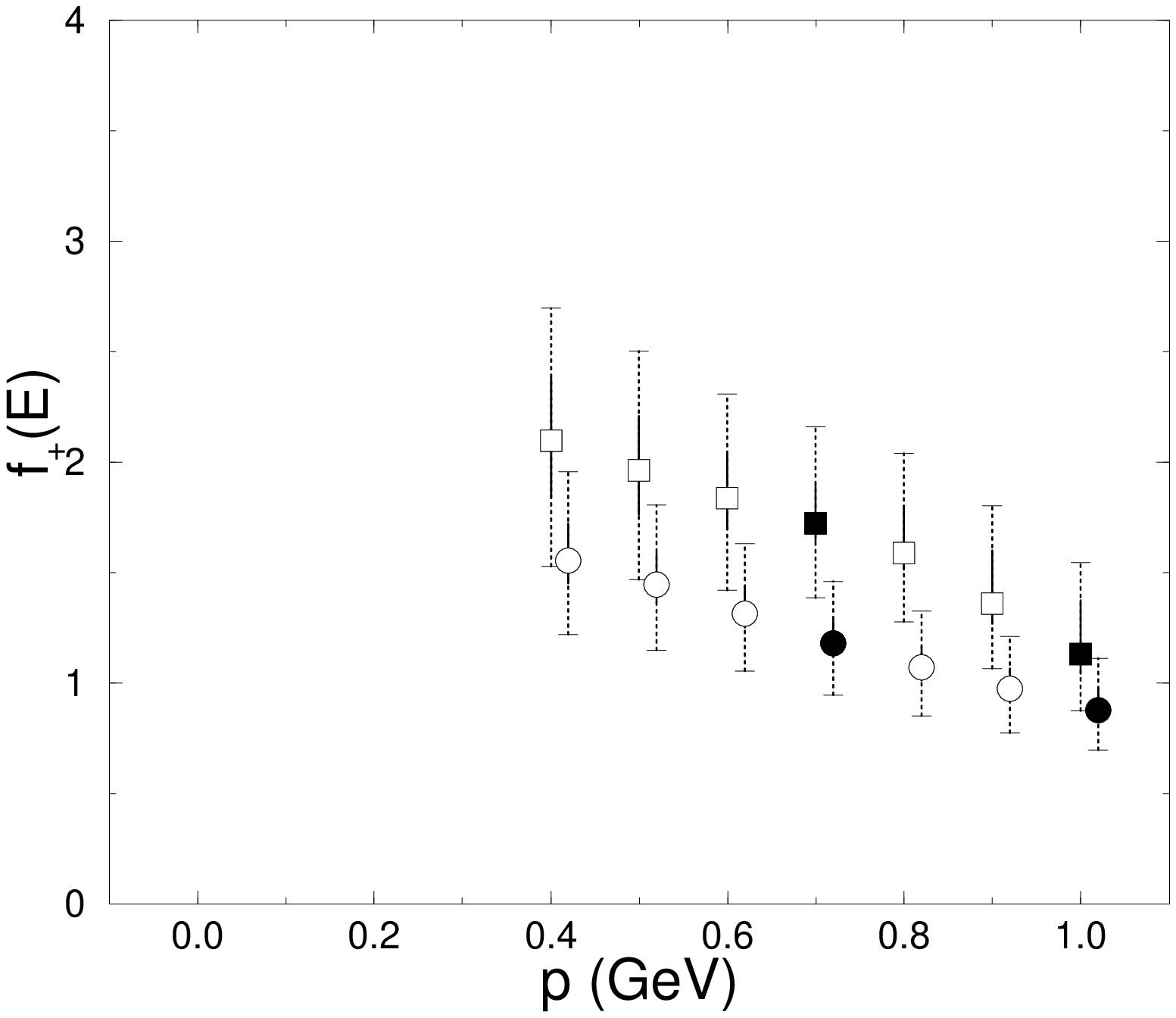}
	\epsfysize=0.45\textheight 
		\epsfbox{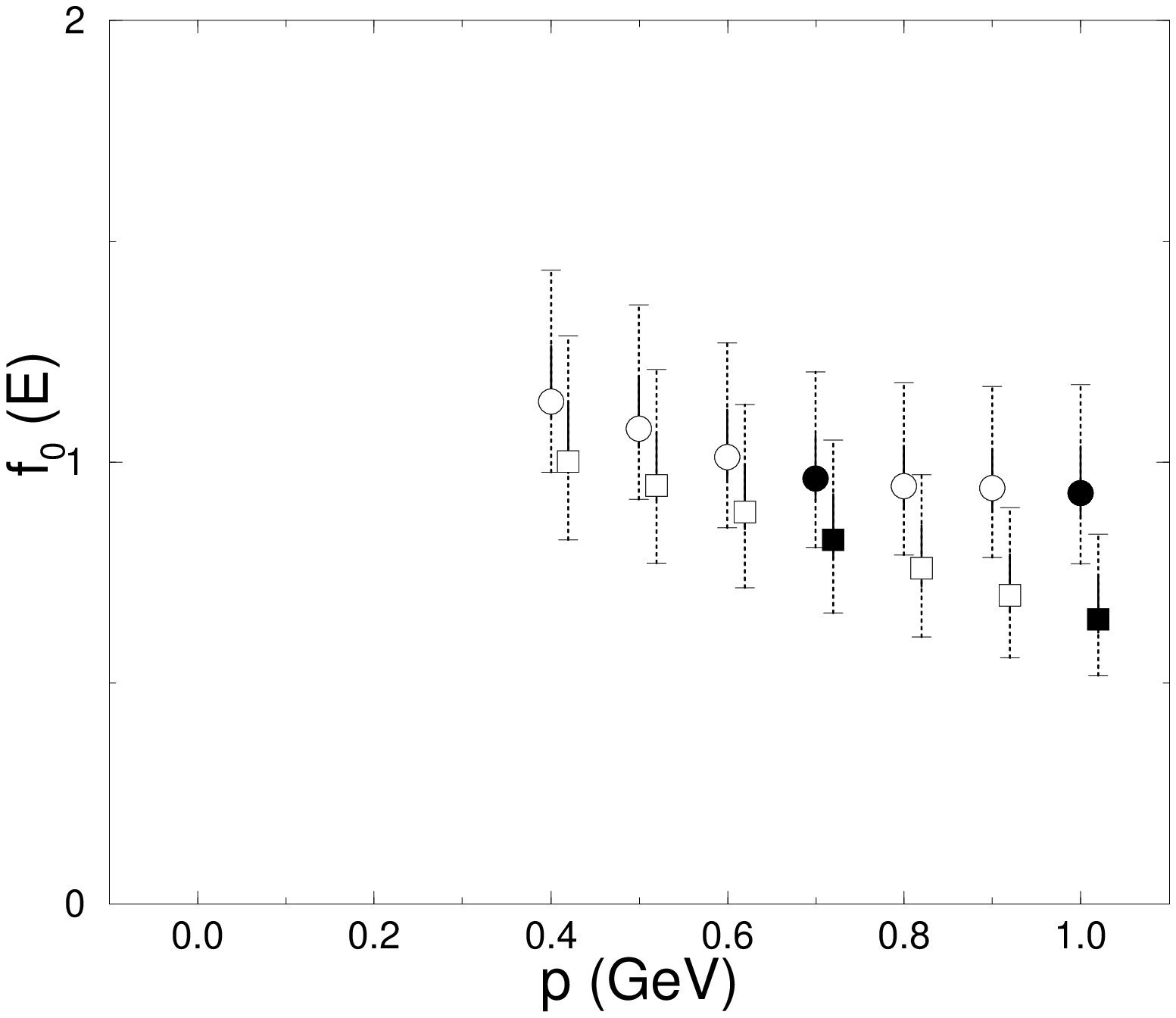}
    \caption[fig:ff]{Momentum dependence of the form factors 
		with all systematic uncertainties included.
		(a) $f_+$ and (b) $f_0$.
		Squares (circles) denote $B$ ($D$) decays.
		Solid symbols are independent of the momentum interpolation.}
	\label{fig:ff}
\end{figure}

The physical form factors $f_+(E)$ and $f_0(E)$ are obtained from
Eqs.~(\ref{f+}) and~(\ref{f0}) using the tabulated 
results for~$f_\perp(E)$ and~$f_\parallel(E)$, physical
meson masses, and energy
$E=\sqrt{m_\pi^2+p^2}$, for each $p=|\bbox{p}_\pi|$ in our set of
pion three-momenta.
They are shown in Fig.~\ref{fig:ff}.
Tables~\ref{tab:B} and~\ref{tab:D} also include the combination
$p^4|f^+|^2/E$;
for massless final-state leptons the differential rates are given by
\begin{eqnarray}
	\frac{d\Gamma_{B\to\pi}}{dp\hfil} = |V_{ub}|^2\,
		(2.9328~\text{ps}^{-1} \text{GeV}^{-4})\,
		\frac{p^4|f_+|^2}{E},
	\label{eq:dGBdp} \\
	\frac{d\Gamma_{D\to\pi}}{dp\hfil} = |V_{cd}|^2\,
		(1.0358~\text{ps}^{-1} \text{GeV}^{-4})\,
		\frac{p^4|f_+|^2}{E}.
	\label{eq:dGDdp} 
\end{eqnarray}
Other differential distributions can be obtained from the latter by
changing variables with $dp/dE=E/p$ and $dp/dq^2=E/2m_Bp$ or~$E/2m_Dp$.
From Eqs.~(\ref{eq:dGBdp}) and~(\ref{eq:dGDdp}) one sees that the
phase-space factor $p^4$ suppresses the rate in the low-momentum
region where we cannot quote~$f_+$.

\section{Comparison with other results}
\label{sec:compare}

In this section we compare our results to recent
published~\cite{Bowler:2000tx,Abada:2000ty} and
preliminary~\cite{Aoki:2000ij} work from lattice QCD.
The comparison is apt, because three different methods for treating the
heavy quark on the lattice have been employed.
We use Wilson fermions with the SW action, normalized to have a
consistent heavy-quark limit.
References~\cite{Bowler:2000tx,Abada:2000ty} use Wilson fermions
(with the SW action and light-quark normalization conditions) at $m_Q$
near and below the charmed quark mass, and extrapolate up to~$m_b$.
Reference~\cite{Aoki:2000ij} uses lattice NRQCD~\cite{Lepage:1987gg} 
(with the power-counting of HQET~\cite{Eichten:1990zv}) and, as we do, 
calculates the form factors directly at the bottom quark mass.

\begin{figure}
	\centering
	\epsfysize=0.5\textheight 
		\epsfbox{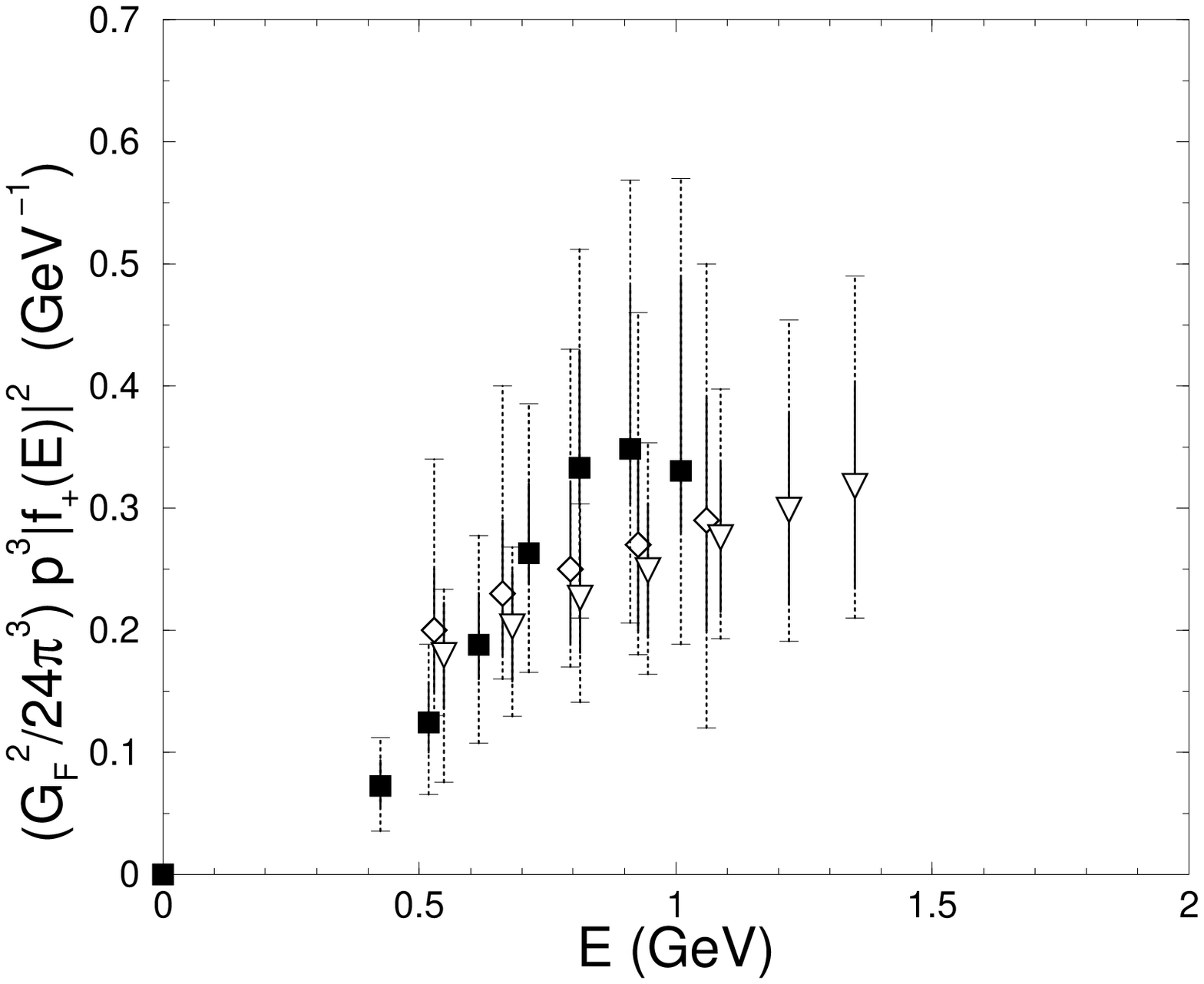}
	\caption[fig:compare]{Comparison of the differential
		decay rate (at fixed lattice spacing) vs.~$E$:
		open diamonds~\cite{Bowler:2000tx}, 
		open triangles~\cite{Abada:2000ty}, and
		solid squares (this work).}
	\label{fig:compare}
\end{figure}

Figure~\ref{fig:compare} shows results from 
Refs.~\cite{Bowler:2000tx,Abada:2000ty} together with ours.
(We do not include results from the JLQCD 
collaboration~\cite{Aoki:2000ij}, because they are still preliminary.
We anticipate that their systematic uncertainties will be similar to 
ours.
At this stage their statistical uncertainties seem surprisingly 
large.)
Within the quoted uncertainties there is broad agreement among the 
three calculations.
There are, however, three noteworthy differences in the analysis of 
the form factors.
These are the lattice spacings at which the calculations are done, the 
procedure for chiral extrapolation, and the treatment of the heavy 
quark.

Our results are based on lattice gauge fields at three lattice 
spacings, given in Table~\ref{tab:sim_details}.
We find that the lattice-spacing dependence of the form factors is 
mild (with our treatment of the heavy quark), even on a relatively 
coarse lattice at $\beta=5.7$.
The results of Refs.~\cite{Bowler:2000tx,Abada:2000ty} are both based 
on only one set of lattice gauge fields (at $\beta=6.2$), whose 
spacing is slightly finer than any of ours.
We believe, therefore, that the lattice spacing effects of the gluons 
and light quarks are not a serious source of error, at the present 
overall level of accuracy, in any of the three works.

In our work, the largest uncertainty comes from the chiral 
extrapolation at fixed pion momentum~$p$.
As explained in Sec.~\ref{subsec:mqp}, this uncertainty arises because 
the linear and quadratic fits give moderately different results.
References~\cite{Bowler:2000tx,Abada:2000ty} do not have enough values 
of the light quark mass to be able to check whether a term quadratic 
in~$m_q$ is needed to describe their data.
Because those works extrapolate at fixed~$q^2$, however, it is 
plausible that the curvature seen in Fig.~\ref{good_extrapsV4} would 
go away, and that a linear chiral extrapolation would then be adequate.

The interpolation in pion momentum, or energy, is another difference.
It leads to the apparent difference in the shape of the spectrum in
Fig.~\ref{fig:compare}.
If we choose the pole form suggested (for small $E$) by
Eqs.~(\ref{soft_parallel}) and~(\ref{soft_perp}), the shape of the
spectrum is less humped, though not as flat as the spectra from
Refs.~\cite{Bowler:2000tx,Abada:2000ty}.
In those works a pole Ansatz different from 
Eqs.~(\ref{soft_parallel}) and~(\ref{soft_perp}) was used.

The most significant difference in the three calculations is the 
treatment of the heavy quark.
Although the same action and similar currents are used, the bare quark 
mass and the normalization of the current are adjusted differently.
The normalization conditions chosen in 
Refs.~\cite{Bowler:2000tx,Abada:2000ty} are designed for the 
$m_Qa\to 0$ limit, and at finite $m_Qa$ they leave systematic 
uncertainties of order~$(m_Qa)^2$.
To reduce these, Refs.~\cite{Bowler:2000tx,Abada:2000ty} calculate 
with the heavy-light meson mass near and below 2~GeV and extrapolate 
up to~$m_B$.
This procedure leads to their largest quoted uncertainty.
The statistical error increases, as it must in any extrapolation.
There are also systematic effects, which are estimated by trying linear
and quadratic fits in~$1/m_Q$.
For at least two reasons, this test may underestimate the systematic 
uncertainty of the extrapolation.
First, the compatibility of the fits shows only that the dependence 
on~$1/m_Q$ is smooth in the employed range of the quark masses.
It does not show that the heavy-quark expansion is reliable below 
2~GeV.
This problem is especially severe for Ref.~\cite{Bowler:2000tx}, 
which has heavy-light meson masses as low as 1.2~GeV.
Second, the lattice artifacts of order~$(m_Qa)^2$ may well be 
amplified by the extrapolation.
This problem would be especially severe for Ref.~\cite{Abada:2000ty}, 
which has $m_Qa$ as high as~0.7.

Our normalization conditions coincide with those above as $m_Qa\to0$.
At finite $m_Qa$, however, they are chosen to eliminate lattice 
artifacts that grow with $m_Qa$~\cite{El-Khadra:1997mp}.
This is made possible by using HQET to match the lattice action and 
current to continuum QCD~\cite{Kronfeld:2000ck}, as reviewed in 
Sec.~\ref{sec:bckgnd}.
The advantage is that, as with lattice 
NRQCD~\cite{Lepage:1987gg,Aoki:2000ij}, the calculations can be done 
directly at $m_Q=m_b$, without an extrapolation in~$1/m_Q$.
Of course, we must assume that HQET is valid for the $b$ quark, but 
that is safer than assuming that it is valid for $m_Q\sim 1$--2~GeV.

A feature of our approach is that it leads to a somewhat complicated 
pattern of heavy-quark discretization errors.
There is, however, a corresponding pattern of systematic uncertainties 
in the results of Refs.~\cite{Bowler:2000tx,Abada:2000ty}.
In particular, there are corrections to the
normalization of order~$\alpha_s (m_Qa)^2$ and of
order~$(m_Qa)^2\Lambda_{\text{QCD}}/m_Q=m_Qa\Lambda_{\text{QCD}}a$
in the $1/m_Q$~term of the heavy-quark expansion.
Estimates of the magnitude of these errors---before or after $1/m_Q$ 
extrapolation---are absent from Refs.~\cite{Bowler:2000tx,Abada:2000ty}.
(The corresponding errors in our work, which we address quantitatively 
in Sec.~\ref{matching}, are of order~$\alpha_s^2$ in the normalization 
and of order~$\alpha_s\Lambda_{\text{QCD}}/m_b$ in the $1/m_Q$~term.)

The calculation of semileptonic form factors for $B\to\pi l\nu$, and
related $D$ decays, has also been carried out with quark models and 
QCD sum rules.
At present the uncertainties from lattice QCD are comparable to those 
based on light-cone sum rules~\cite{Ball:1998tj,Khodjamirian:2000ds}.
The latter have the advantage that they are most applicable for 
energetic pions, so (for $B$ decay) they overlap better with th 
distribution of events in an experiment.
On the other hand, it seems difficult to reduce the uncertainties from 
sum rules down to the level of a few percent, which will be needed to 
match the precision of the $B$ factories. 
As discussed in the following section, however, all uncertainties of 
the form factors are reducible with lattice QCD.

\section{Conclusions}
\label{sec:conclude}

In this paper we have presented results for the form factors and
differential decay rates for the semileptonic decays $B\to\pi l\nu$ 
and $D\to\pi l\nu$.
The total uncertainties are 30--35~percent (for the rate) and, 
hence, would yield a theoretical uncertainty to the CKM matrix 
of 15--18~percent.
We have attempted a complete analysis of the systematic uncertainties, 
at least within the quenched approximation.
A rough estimate of the additional error from quenching is 
10--20~percent (on the rate).

A more important, thought less specific, result of this paper is a 
demonstration that, within the quenched approximation, all 
uncertainties are controllable.
Table~\ref{tab:cpu} gives a sketch of what is needed to reduce all 
sources of uncertainty.
\begin{table}
	\centering
	\caption{Strategies for reducing statistical and systematic uncertainties
		of semileptonic form factors.}
	\label{tab:cpu}
	\begin{tabular}{ll}
		uncertainty         & \multicolumn{1}{c}{strategy} \\ 
		\hline
		statistical         & more lattice gauge fields;
			better $m_q$ extrapolation \\
		excited states      & longer time extent; better operators \\
		$\bbox{p}$ extrapolation & larger finite volume; better statistics  \\
		$m_q$ extrapolation & better statistics; more values of $m_q$;
		  fixed-$E$ extrapolation \\
		adjusting $m_Q$     & unquench  \\
		HQET matching       & match up to $\alpha_s\Lambda_{\text{QCD}}/m_Q$,
			$(\Lambda_{\text{QCD}}/m_Q)^2$ \\
		$a$ dependence      & more lattices \\
		definition of $a$   & unquench \\
		quenching           & unquench \\
	\end{tabular}
\end{table}

In almost every case, the remedy is simply more computer time.
That, in fact, is promising, since the computer used in this work is 
already ten years old.
Given the experience of the 
CP-PACS~\cite{AliKhan:2000eg,AliKhan:2000zr} and 
MILC~\cite{Bernard:2000nv} collaborations with heavy-light decay 
constants, it should be feasible to repeat our analysis on a modern 
supercomputer with unquenched gauge fields.
In summary, there do not appear to be any technical roadblocks to 
reducing the uncertainties in lattice calculations to a few percent or 
better, over the course of the present round of experiments.

In the case of the uncertainty labeled ``HQET matching'' better 
calculations of the various short-distance coefficients introduced in 
Sec.~\ref{sec:bckgnd} will be needed to be sure that the total 
uncertainty is only a few percent.
This is not a computational problem but a theoretical one, which 
arises also in calculations with lattice NRQCD.
The alternative would be to reduce the lattice spacing dramatically, 
so that $m_Qa$ and $\Lambda_{\text{QCD}}/m_Q$ can be simultaneously 
small.
But, since computer requirements scale as $a^{-1}$ to a high power, 
that would seem to be a long way off.
One would also have to sacrifice some other improvements, such 
as removing the quenched approximation.

For semileptonic decays of $B$ or$D$ to vector mesons such as
$\rho$ and $\omega$ more study is needed.
The vector mesons decay hadronically, and
in the quenched approximation these decays are absent.
Even in unquenched calculations, however, there are still issues 
that may need explicit analysis.
In particular, if the calculations are done at largish light quark 
masses, the decay may be kinematically forbidden.
Because there is not yet much experience with unquenched calculations, 
it is not yet clear whether one can smoothly extrapolate vector meson 
properties from this region to the physical meson masses.
It is, thus, hard to anticipate how well lattice QCD will do here.
This is unfortunate, because the experimental errors for semileptonic 
decays into vector mesons are expected to be somewhat smaller.

In any case, semileptonic decays of $B$ mesons ultimately will provide 
one of the most accurate constraints on the unitarity triangle, 
through a determination of~$|V_{ub}|$.
Indeed, if new physics lurks behind $B^0$-$\bar{B}^0$ mixing, it is 
essential to have constraints on the CKM matrix through charged-current
interactions like $b\to c$ and $b\to u$.

\acknowledgments
High-performance computing was carried out on ACPMAPS; we thank past
and present members of Fermilab's Computing Division for designing,
building, operating, and maintaining this supercomputer, thus making
this work possible.
Fermilab is operated by Universities Research Association Inc.,
under contract with the U.S.\ Department of Energy.
AXK is supported in part by the DOE OJI program under contract
DE-FG02-91ER40677 and through the Alfred P. Sloan Foundation.
SMR would like to thank the Fermilab's Theoretical Physics Department
for hospitality while part of this work was being carried out.

\end{document}